\documentclass{article}

\usepackage{fullpage}
\usepackage{amsmath,amsfonts}
\usepackage{booktabs}
\usepackage{longtable}

\usepackage{caption}   
\usepackage{array}
\usepackage{xurl}
\usepackage{pdfpages}
\usepackage{hyperref}
\usepackage{nameref}
\usepackage{authblk}

\usepackage{tikz}
\usetikzlibrary{backgrounds}
\usetikzlibrary{calc}
\usetikzlibrary{shapes.geometric}
\definecolor{grey}{rgb}{0.5,0.5,0.5}
\definecolor{darkgrey}{rgb}{0.2,0.2,0.2}
\definecolor{verylightgrey}{rgb}{0.9,0.9,0.9}
\definecolor{lightgrey}{rgb}{0.8,0.8,0.8}
\definecolor{grey7}{rgb}{0.966,0.966,0.966}
\definecolor{grey6}{rgb}{0.933,0.933,0.933}
\definecolor{grey5}{rgb}{0.9,0.9,0.9}
\definecolor{grey4}{rgb}{0.866,0.866,0.866}
\definecolor{grey3}{rgb}{0.833,0.833,0.833}
\definecolor{grey2}{rgb}{0.8,0.8,0.8}
\definecolor{grey1}{rgb}{0.766,0.766,0.766}

\usepackage{natbib}
\usepackage{url}

\bibpunct{(}{)}{;}{a}{}{,}

\title{Do Network Mechanisms in Science Vary Across Research Topics? Contextual Relational Hyperevent Models for Large-Scale Bibliometric Data}

\author[1]{J\"urgen Lerner}
\author[2]{Mark Wittek\thanks{Corresponding author: \href{mailto:wittekm@ceu.edu}{wittekm@ceu.edu}}}
\author[3]{Raphael H. Heiberger}
\affil[1]{Department of Computer and Information Science, University of Konstanz, Konstanz, Germany}
\affil[2]{Department of Network and Data Science, Central European University, Vienna, Austria}
\affil[3]{Institute for Social Sciences, University of Stuttgart, Stuttgart, Germany}
\date{\small ORCID iDs: J.~Lerner \href{https://orcid.org/0000-0001-6565-1873}{0000-0001-6565-1873}; M.~Wittek \href{https://orcid.org/0000-0003-3818-1997}{0000-0003-3818-1997}; R.~H.~Heiberger \href{https://orcid.org/0000-0003-3465-7214}{0000-0003-3465-7214}}

\begin{document}

\maketitle

\clearpage

\begin{abstract}
\noindent Collaboration and citation networks in science are shaped by well-documented mechanisms such as preferential attachment, gender homophily, and self-citation. Whether these mechanisms operate uniformly across research areas, however, is rarely tested, because statistical network models that estimate several mechanisms jointly are difficult to apply to large bibliometric data and to compare across subnetworks. We introduce contextual relational hyperevent models (C-RHEMs), which extend relational hyperevent models by allowing network effects to vary across research topics within a single model. C-RHEMs treat each publication as one event linking its authors to its references, accommodate reference lists spanning several topics, and scale to hundreds of thousands of publications on standard hardware. In a second step, meta-analysis and meta-regression quantify heterogeneity in topic-specific effects and relate it to topic characteristics. We illustrate the approach with about 590,000 oncology publications from the PubMed Knowledge Graph, assigned to 32 OpenAlex research topics. We find that preferential attachment, gender homophily, and self-citation vary substantially across topics, with between-topic standard deviations of 30 to 70 percent of the corresponding pooled effects. Topic age, size, and gender composition account for between a third and two-thirds of this variation: older topics exhibit stronger preferential attachment, self-citation, and gender homophily, and topics with a higher share of male authors exhibit stronger self-citation. The results caution against treating network mechanisms estimated on pooled data as uniform, and C-RHEMs offer a scalable way to study how they differ across research areas.

\end{abstract}

\clearpage

\section{Introduction}
\label{sec:intro}

Collaboration and citation networks are widely studied in the science of science, sociology of science, and scientometrics to understand how networks shape innovation or facilitate the success of individuals and teams. While networks are often treated as a cause for an outcome of interest, a long line of research also investigates the mechanisms that underpin network formation: highly cited papers attract further citations and productive scientists attract further collaborators \citep{merton_matthew_1968, price_general_1976, newman_scientific_2001, eom_characterizing_2011}, scientists tend to collaborate with and cite others of the same gender \citep{main_gender_2014, potthoff_is_2017}, and authors build on their own prior work through self-citation \citep{wolfgang_bibliometric_2004, azoulay_self-citation_2020}. Whereas most of this work assumes that network mechanisms operate similarly across domains, multiple studies have pointed to crucial differences in publication and citation practices across research areas. For instance, the interpretation of citation indicators should take into account disciplinary differences in citation behavior \citep{waltman_review_2016} and similar differences have been documented for self-citation \citep{aksnes_macro_2003} and for collaboration patterns \citep{kronegger_collaboration_2012}. However, much less is known about whether the \emph{mechanisms} generating collaboration and citation networks differ across research areas, how large such differences are, and which characteristics of research areas account for them. This paper introduces a method for answering these questions at the scale of contemporary bibliometric data.\footnote{The data and analysis code are publicly available (cf.\ \nameref{para:data}).}

Several accounts from the social sciences highlight that research areas are marked by different forms of social organization. Crane's account of invisible colleges illustrates that the ways in which scientists collaborate and cite each other depend on the developmental stage of their research communities, which evolve from small, emerging groups into consolidated paradigms characterized by specialization and the centralization of recognition and resources \citep{crane_invisible_1972}. Similarly, Knorr Cetina's work on epistemic cultures shows that the social organization of research differs with its material and organizational conditions, and can range from large collaborations built around shared infrastructures to small laboratory teams \citep{cetina_epistemic_1999}. These perspectives suggest that research areas differ not only in their substantive focus but also in the mechanisms through which networks inside them emerge and evolve.

Network ecology \citep{mcfarland_network_2014} provides an explicit framework for theorizing this variation by explaining how contextual characteristics---such as age, size, or demographic composition---shape the mechanisms underlying social networks. Drawing on an ecological analogy, it posits that network relations respond to contexts, much as organisms adapt to ecological niches. \citet{mcfarland_network_2014} introduced the framework to study adolescents' friendship networks, showing that mechanisms such as triadic closure and gender homophily vary across school environments. For example, friendships in larger schools exhibit greater reciprocity and transitivity, which the authors attribute to students' stronger demand for trust and support in larger, more anonymous settings. Although subsequent studies have further developed this perspective \citep{simpson_farm_2019, doehne_network_2024}, applications to science remain rare (for exceptions, see \citet{kim2024networking}). While the contribution of our article is methodological, we use network ecology as an orienting framework informing the design of our analysis: we examine the three contextual characteristics it emphasizes---age, size, and demographic composition---and we follow its practice of comparing network models across contexts by means of meta-analysis.

Here, we build on a small number of studies that have examined variation in network mechanisms for scientific networks. Using stochastic actor-oriented models, \citet{kronegger_collaboration_2012} and \citet{ferligoj_scientific_2015} estimated the mechanisms driving co-authorship networks separately for the scientific fields and disciplines of the Slovenian research system. They found that preferential attachment operates differently across disciplines and cannot be reduced to a single global mechanism, and conclude that disciplines are affected differently by local institutional arrangements and publishing cultures. Research on gender homophily has similarly compared collaboration and citation patterns across fields, finding that the strength and consequences of homophily differ with the gender composition of a field and between cohorts of scientists \citep{zhou_gender_2024, santos_homophily_2024}. These studies suggest that network mechanisms in science vary across research areas. Because the models are estimated separately for each area, however, comparing their parameters is not straightforward; \citet{ferligoj_scientific_2015}, for instance, transform their estimates into bounded importance measures and cluster disciplines on these measures, since the raw estimates are not directly comparable across disciplines of different sizes. Extending this line of research to large bibliometric datasets, and to a formal assessment of which characteristics of research areas explain the variation, raises two methodological challenges.

The first challenge concerns scale and model degeneracy. Statistical network models that estimate several mechanisms simultaneously, most prominently (temporal) exponential random graph models \citep{lusher2013exponential, krivitsky2014separable} and stochastic actor-oriented models \citep{snijders2005models}, rely on repeated simulation of complete networks. Estimation therefore becomes computationally difficult or infeasible for networks of hundreds of thousands of nodes, and degeneracy and convergence problems remain persistent challenges. Applications of these models to scientific networks have accordingly focused on comparatively small networks, such as national research systems. Models that do scale to large bibliometric data, such as configuration models and their variants \citep{newman2001random, uzzi2013atypical}, preserve node degrees while randomizing other structures. Yet, they do not estimate the simultaneous effects of several mechanisms, and they treat mechanisms such as preferential attachment as invariant rather than allowing their strength to vary across subnetworks.

The second challenge concerns the structure of bibliometric data. Publications are events that occur at specific points in time and that are more naturally represented as multivariate counting processes than as relational states \citep{perry2013point, lerner2025relational}. Moreover, each publication simultaneously links a set of authors to the paper and the paper to a set of references. Reducing such hyperedges to dyadic ties---for example, by connecting every pair of coauthors---leads to information loss, inflates the number of observations, and introduces structural artifacts such as an overabundance of closed triangles \citep{battiston2020networks, chodrow2020annotated, lerner2023micro}. A single paper with 100 authors, for instance, produces 4,950 coauthor ties and 161,700 closed triangles.

Relational hyperevent models (RHEMs) address both challenges \citep{lerner2023relational, lerner2025relational}. Building on relational event models \citep{butts2008relational, bianchi2024relational}, RHEMs model events involving arbitrary numbers of nodes directly, treating each publication as a single observation that links its authors to its references. They are estimated by partial likelihood with sampled risk sets, which keeps estimation feasible for networks with more than a million nodes on standard desktop hardware \citep{lerner2020reliability, lerner2025relational}, and they have recently been extended to tripartite events that additionally link publications to keywords \citep{fabbrucci_tripartite_2026}. Existing RHEMs, however, estimate a single set of parameters for the entire network and thus assume that network mechanisms operate uniformly across its parts.

We relax this assumption by introducing \emph{contextual} relational hyperevent models (C-RHEMs). C-RHEMs estimate topic-specific intercepts and topic-specific slopes for selected network effects within a single partial-likelihood model, so that the resulting estimates are directly comparable across topics. Because reference lists frequently span several topics, topic membership enters the model as the share of a paper's references belonging to each topic, which accommodates overlapping topical subnetworks. In a second step, we quantify the heterogeneity of the topic-specific estimates with standard meta-analytic statistics and relate it to characteristics of the topics through meta-regression, an approach established for comparing network models across contexts \citep{an_multilevel_2015, smith_ethnic_2016}.

We illustrate C-RHEMs with the co-evolving collaboration and citation networks of about 590,000 oncology publications from the PubMed Knowledge Graph \citep{xu_pubmed_2025}, assigned to 32 research topics of the OpenAlex classification \citep{priem_openalex_2022}. Oncology provides a useful test case because its topics share a common scientific objective while differing substantially in age, size, and gender composition \citep{maman_history_2018, olson_history_1989}. We estimate topic-specific variation in three mechanisms: preferential attachment in collaboration and in citation, gender homophily in collaboration and in citation, and self-citation. We then examine whether this variation is associated with three characteristics of research topics that prior work links to network processes: topic size, which has been associated with inequality and slower turnover of ideas \citep{chu_slowed_2021, linzhuo_social_2020, wittek_development_2023}; gender composition \citep{kim_gendered_2022}; and topic age, following Crane's developmental account \citep{crane_invisible_1972}.

We find that all three mechanisms vary substantially across topics: between-topic standard deviations amount to 30 to 70 percent of the corresponding pooled effects, and almost all of the observed dispersion reflects genuine differences between topics rather than estimation uncertainty. Topic age, size, and gender composition jointly account for between a third and two-thirds of this variation. Older topics show stronger preferential attachment, self-citation, and gender homophily; topics with a higher share of male authors show stronger self-citation; and topic size shows few systematic associations. These results suggest that network mechanisms estimated on pooled data can mask substantial heterogeneity across research areas, so that conclusions drawn from one area need not carry over to others. C-RHEMs provide a scalable way to estimate this heterogeneity and to investigate its sources.

The remainder of the paper is organized as follows. Section~\ref{sec:models} introduces RHEMs for coauthoring and citation networks, develops the contextual extension, and describes the meta-analytic second stage. Section~\ref{sec:data} describes the data, Section~4 reports the results, and Section~5 discusses implications and limitations.

\section{Extending Relational Hyperevent Modeling}
\label{sec:models}

\subsection{RHEM for coauthoring and citation networks}

RHEMs are in the tradition of relational event models \citep[REM,][]{butts2008relational}, a family of models that specify the intensity, hazard, or rate of interaction events in networks. Originally formulated for events connecting a sender with a receiver, such as a phone call or message, REMs have been generalized to events involving a whole set of receivers \citep{perry2013point,lerner2023relational} or to events on \emph{two-mode hyperedges}, connecting the set of authors of a scientific publication to its list of references \citep{lerner2025relational}. 

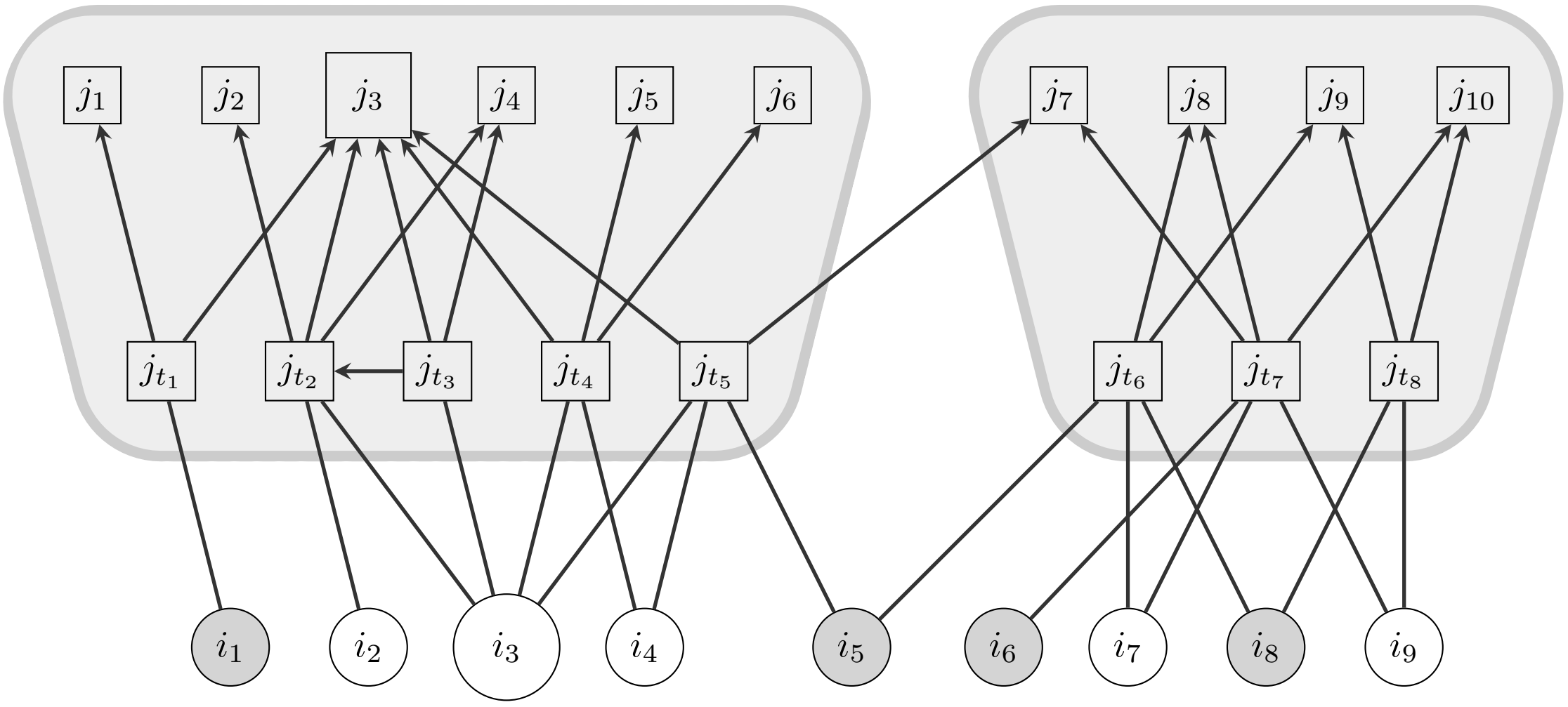
\begin{figure}[!ht]
  \begin{center}
    \begin{tikzpicture}[scale=1.3,>=stealth]
      \tikzstyle{actor}=[circle,minimum size=1mm,draw=black];
      \tikzstyle{field}=[diamond,minimum size=1mm];
      \tikzstyle{work}=[rectangle,minimum size=1mm,draw=black];
      \node[work] (w1) at (-8.0,2.0) {$j_1$};
      \node[work] (w2) at (-7.0,2.0) {$j_2$};
      \node[work,minimum size=8mm] (w3) at (-6.0,2.0) {$j_3$};
      \node[work] (w4) at (-5.0,2.0) {$j_4$};
      \node[work] (w5) at (-4.0,2.0) {$j_5$};
      \node[work] (w6) at (-3.0,2.0) {$j_6$};
      \node[work] (w7) at (-1.0,2.0) {$j_7$};
      \node[work] (w8) at (0.0,2.0) {$j_8$};
      \node[work] (w9) at (1.0,2.0) {$j_9$};
      \node[work] (w10) at (2.0,2.0) {$j_{10}$};
      
      \node[work] (wt1) at (-7.5,-0.0) {$j_{t_1}$};
      \node[work] (wt2) at (-6.5,-0.0) {$j_{t_2}$};
      \node[work] (wt3) at (-5.5,-0.0) {$j_{t_3}$};
      \node[work] (wt4) at (-4.5,-0.0) {$j_{t_4}$};
      \node[work] (wt5) at (-3.5,-0.0) {$j_{t_5}$};
      \node[work] (wt6) at (-0.5,-0.0) {$j_{t_6}$};
      \node[work] (wt7) at (0.5,-0.0) {$j_{t_7}$};
      \node[work] (wt8) at (1.5,-0.0) {$j_{t_8}$};

      \draw[color=darkgrey,line width=1pt,->] (wt1) to (w1);
      \draw[color=darkgrey,line width=1pt,->] (wt1) to (w3);
      \draw[color=darkgrey,line width=1pt,->] (wt2) to (w3);
      \draw[color=darkgrey,line width=1pt,->] (wt2) to (w2);
      \draw[color=darkgrey,line width=1pt,->] (wt2) to (w4);
      \draw[color=darkgrey,line width=1pt,->] (wt3) to (w3);
      \draw[color=darkgrey,line width=1pt,->] (wt3) to (w4);
      \draw[color=darkgrey,line width=1pt,->] (wt4) to (w3);
      \draw[color=darkgrey,line width=1pt,->] (wt4) to (w5);
      \draw[color=darkgrey,line width=1pt,->] (wt4) to (w6);
      \draw[color=darkgrey,line width=1pt,->] (wt5) to (w3);
      
      \draw[color=darkgrey,line width=1pt,->] (wt5) to (w7);
      \draw[color=darkgrey,line width=1pt,->] (wt3) to (wt2);

      \draw[color=darkgrey,line width=1pt,->] (wt6) to (w8);
      \draw[color=darkgrey,line width=1pt,->] (wt6) to (w9);
      \draw[color=darkgrey,line width=1pt,->] (wt7) to (w7);
      \draw[color=darkgrey,line width=1pt,->] (wt7) to (w8);
      \draw[color=darkgrey,line width=1pt,->] (wt7) to (w10);
      \draw[color=darkgrey,line width=1pt,->] (wt8) to (w9);
      \draw[color=darkgrey,line width=1pt,->] (wt8) to (w10);

      \node[actor,fill=grey3] (a1) at (-7,-2.0) {$i_1$};
      \node[actor] (a2) at (-6,-2.0) {$i_2$};
      \node[actor,minimum size=10mm] (a3) at (-5,-2.0) {$i_3$};
      \node[actor] (a4) at (-4,-2.0) {$i_4$};
      \node[actor,fill=grey3] (a5) at (-2.5,-2.0) {$i_5$};
      \node[actor,fill=grey3] (a6) at (-1.4,-2.0) {$i_6$};
      \node[actor] (a7) at (-0.5,-2.0) {$i_7$};
      \node[actor,fill=grey3] (a8) at (0.5,-2.0) {$i_8$};
      \node[actor] (a9) at (1.5,-2.0) {$i_9$};

      \draw[color=darkgrey,line width=1pt] (a1) to (wt1);
      \draw[color=darkgrey,line width=1pt] (a2) to (wt2);      
      \draw[color=darkgrey,line width=1pt] (a3) to (wt2);
      \draw[color=darkgrey,line width=1pt] (a3) to (wt3);
      \draw[color=darkgrey,line width=1pt] (a3) to (wt4);
      \draw[color=darkgrey,line width=1pt] (a3) to (wt5);
      \draw[color=darkgrey,line width=1pt] (a4) to (wt4);
      \draw[color=darkgrey,line width=1pt] (a4) to (wt5);

      \draw[color=darkgrey,line width=1pt] (a5) to (wt5);
      \draw[color=darkgrey,line width=1pt] (a5) to (wt6);
      \draw[color=darkgrey,line width=1pt] (a6) to (wt7);
      \draw[color=darkgrey,line width=1pt] (a7) to (wt6);
      \draw[color=darkgrey,line width=1pt] (a7) to (wt7);
      \draw[color=darkgrey,line width=1pt] (a8) to (wt6);
      \draw[color=darkgrey,line width=1pt] (a8) to (wt8);
      \draw[color=darkgrey,line width=1pt] (a9) to (wt7);
      \draw[color=darkgrey,line width=1pt] (a9) to (wt8);

      \begin{pgfonlayer}{background}
        \foreach \nodename in {w1,w2,w3,w4,w5,w6,w7,w8,w9,w10,wt1,wt2,wt3,wt4,wt5,wt6,wt7,wt8} {
          \coordinate (\nodename') at (\nodename);
        }
        \path[draw=grey2,line width=1.7cm,dashed, line cap=round, line join=round] 
        (w2') to (w3') to (w4') to (w5') to (w6') to (wt5') to (wt4') to (wt3') to (wt2') to (wt1') to (w1') to (w2') -- cycle;
        \path[fill=grey6,draw=grey6,line width=1.5cm, line cap=round, line join=round] 
        (w1') to (w2') to (w3') to (w4') to (w5') to (w6') to (wt5') to (wt4') to (wt3') to (wt2') to (wt1') to (w1') -- cycle;
        \path[draw=grey2,line width=1.7cm, line cap=round, line join=round] 
        (w7') to (w8') to (w9') to (w10') to (wt8') to (wt7') to (wt6') to (w7') -- cycle;
        \path[fill=grey6,draw=grey6,line width=1.5cm, line cap=round, line join=round] 
        (w7') to (w8') to (w9') to (w10') to (wt8') to (wt7') to (wt6') to (w7') -- cycle;
      \end{pgfonlayer}
    \end{tikzpicture}
  \end{center}
\caption{\textbf{Stylized example of the publication events of papers $j_{t_1},\dots,j_{t_8}$}. Papers (rectangles) are connected to their authors (circles) and to their references (rectangles). 
Papers belong to one of two scientific topics, indicated by the gray-shaded areas.
Authors have a gender variable, indicated by the darkness of their circles. The larger topic on the left-hand side gives rise to stronger preferential attachment in citations (e.\,g., paper $j_3$ attracts more citations than average) and stronger preferential attachment in authorship (e.\,g., author $i_3$ published more than average). This topic also gives rise to stronger gender-homophily in co-authoring, since most co-author teams are almost perfectly gender-homogeneous. The smaller topic on the right-hand side features more equally distributed citation counts and authorship counts and has gender-mixed author teams.
}
\label{fig:stylized}
\end{figure}

In this paper, we extend RHEM for paper publication events, given in the form
\[
j=(t,\ell,I,J)\enspace,
\]
indicating the publication of paper $j$ in the scientific context (or topic) labeled by $\ell$, where $j$ is written by authors $I$, cites the references $J$, and is published at time $t$. Taking such a multi-mode hyperedge as the unit of observation facilitates achieving three objectives. First, jointly considering authors and references allows us to tackle our core network mechanisms---some concerning team formation (preferential attachment and homophily in the coauthor network), some concerning citation decisions (preferential attachment and homophily in the citation network), and yet others concerning relations between authors and papers (self-citation)---simultaneously in a unified model \citep{burgdorf_communities_2024,lerner2025relational}. Second, bundling the entire sets of authors and references in a single observation allows for testing and controlling for \emph{hyperedge effects} \citep{lerner2023relational}, such as the tendency of groups of scientists to repeatedly co-author papers or the tendency of sets of papers to be frequently cited together. It has been demonstrated that such hyperedge effects can be very strong in publication networks \citep{lerner2025relational}. Third, labeling publications by their scientific topic allows us to assess contextual variation of effects, in line with our core objectives. 
Figure~\ref{fig:stylized} presents a stylized example illustrating patterns of citation and authorship over two scientific topics and hints at contextual variation of various network effects.

Network effects explaining such publication events are operationalized by a vector of \emph{hyperedge covariates} $x(t,\ell,I,J)\in\mathbb{R}^p$ which broadly fall into two categories: covariates dependent on exogenous attributes and \emph{endogenous} effects, dependent on the history of the process, that is, on prior publications $j'=(t',\ell',I',J')$ with $t'<t$. Coauthor homophily with respect to gender is an example of an exogenous effect as it measures the homogeneity of the ``gender'' variable (an attribute of scientists which is exogenous to the publication process) among the coauthor team $I$. Preferential attachment in the citation network is an example of an endogenous effect as it depends on the number of prior publications that previously cited one or several of the references $J$ of the focal paper. In our models, we include not only covariates that operationalize the network mechanisms of core interest but also control effects, which often formalize intuitively plausible and empirically strong alternative explanations for scientific collaboration and references. These control effects include the tendency to cite papers within the same topic, the tendency of scientists to repeatedly cite the same references, or the tendency to cite papers together with some of their references. While the substantive scientific interest in some of these (almost ``obvious'') effects might be limited, their explanatory power \citep{lerner2025relational} entails the need to control for them---since otherwise unexplained dependencies might distort effects of core interest.

The response variable of RHEM considered here is the \emph{intensity}, or \emph{hazard rate} \citep{aalen2008survival,perry2013point} $\lambda(t,\ell,I,J)$, which can be understood as the expected number of events per time unit, or alternatively as the instantaneous probability of observing an event at $t$ \citep{butts2008relational}, on the focal instance---in our case, observing a publication in the scientific topic $\ell$, written by authors $I$ and citing references $J$. In the framework of Cox regression models \citep{cox1972regression,aalen2008survival,perry2013point}, the \emph{relative intensity} $\lambda_1$ is typically specified as a parametric function of the covariates as follows:
\begin{equation}
\label{eq:intensity}
\lambda_1(t,\ell,I,J,\beta)=\exp\left[\beta^T\cdot x(t,\ell,I,J)\right]=\exp\left[\sum_{h=1}^p\beta_h\cdot x_h(t,\ell,I,J)\right]\enspace.    
\end{equation}
The relative intensity, or risk factor, is the factor by which the intensity on the focal hyperedge $(I,J)$ is higher or lower than the \emph{baseline intensity}, that is, the expected number of publications on a "typical", or average hyperedge. Our research interest, indeed, lies in these relative factors, rather than in the baseline intensity. For example, we analyze whether a publication with a set of authors $I$ is more likely than one with authors $I'$ if $I$ has a more homogeneous gender distribution. Likewise, we are interested in whether a set of papers $J$ is more likely to be the reference list of a novel paper if the papers in $J$ have previously been cited more often. In contrast, we are less interested in how the average publication frequency over all possible author teams or reference list changes over time. We note in passing that there are methods for estimating the baseline intensity in Cox regression models \citep{aalen2008survival}. Equation (\ref{eq:intensity}) specifies the relative intensity in log-linear form. If a parameter $\beta_h$ is positive (negative), then increasing the value of the associated covariate $x_h(t,\ell,I,J)$ tends to increase (decrease) the relative intensity.

RHEM are estimated based on given empirical data representing a sequence of publication events:
\[
j_1=(t_1,\ell_1,I_1,J_1),\dots,j_n=(t_n,\ell_n,I_n,J_n)\enspace.
\]
The paper $j_m$ is published at time $t_m$ and is labeled by the topic $\ell_m\in\mathcal{L}$, where $\mathcal{L}$ is the entire set of labels. Its authors $I_m\subseteq\mathcal{I}_{t_m}$ are a subset of the scientists $\mathcal{I}_{t_m}$ that could possibly publish at $t_m$ and its references $J_m\subseteq\mathcal{J}_{t_m}$ are a subset of the papers $\mathcal{J}_{t_m}$ that could possibly be cited at $t_m$, that is, prior work. Given such data, RHEMs specify the partial likelihood within the framework of Cox regression models \citep{cox1972regression,aalen2008survival} by
\begin{equation}
\label{eq:likelihood}
L(\beta)=\prod_{m=1}^n\frac{\lambda_1(t_m,\ell_m,I_m,J_m,\beta)}{\sum_{(I,J)\in\mathcal{R}(t_m)}\lambda_1(t_m,\ell_m,I,J,\beta)}
\enspace.
\end{equation}
In the equation above, $\mathcal{R}(t_m)$ is the (sampled) \emph{risk set} comprising alternative combinations $(I,J)$ of authors $I\subseteq\mathcal{I}_{t_m}$ and sets of papers $J\subseteq\mathcal{J}_{t_m}$ that could have been the authors and references of the paper $j_m$, instead of the actual authors $I_m$ and references $J_m$. The partial likelihood above is equivalent to a conditional multinomial logit model, specifying a discrete choice model that selects the true authors and references out of a sampled space of alternatives $\mathcal{R}(t_m)$. We condition the risk set $\mathcal{R}(t_m)$ on sets of authors $I$ and references $J$ that have the same size as the observed author set $I_m$ and reference list $J_m$. This defines a size-stratified RHEM \citep{lerner2023relational}, where the baseline intensity can vary by hyperedge size (and not only over time). Since the entire space of alternatives is prohibitively large, some form of sampling is necessary. In the empirical analysis of this paper, we apply case-control sampling \citep{borgan1995methods}, as it has been suggested for RHEM \citep{lerner2023relational,lerner2025relational}.
Given the hyperedge covariates $x$ for all instances in the sampled risk set, parameters are estimated by maximizing the partial likelihood (\ref{eq:likelihood}), which can be done with standard statistical software, for example with the \texttt{coxph} function in the \texttt{R} package \texttt{survival} \citep{therneau2000modeling,therneau2024package}. Time is measured in years, so that publications from the same year with the same number of authors and references share a risk set. We provide more information on covariate computation, sampling, and parameter estimation below.

\subsection{Contextual variation of effects in RHEM}
\label{sec:contextual}

Going beyond the models by \citet{lerner2025relational,espinosa2025socio}, in this
article, we propose an extension of RHEM to account for contextual variation over
scientific topics.
We estimate topic-specific intercepts (indicating whether papers published in certain
topics are more or less likely to be cited) and topic-specific slopes (indicating the
variation of the five core effects across topics)\footnote{While we theoretically
consider three network mechanisms (preferential attachment, self-citation, and gender
homophily), operationalizing them requires five network effects: two capturing
preferential attachment in collaboration and citation networks, one for self-citation,
and two for gender homophily in collaboration and citation networks.} as fixed
parameters---that is, as separate coefficients per topic---rather than as random effects
drawn from a common distribution. We therefore do not impose partial pooling across
topics but estimate each topic's effects from those events that have a non-zero
membership to the focal topic (compare the definition of the topic-membership vector
$L(J)$ below). Estimating fixed parameters, rather than random effects, is possible for
this study since even the smallest topics considered in our analysis contain more than a
thousand publications and thus provide enough information for estimating parameters. In
other data that contains topics, or other indicators of context, with a much smaller
number of observations, it likely becomes necessary to impose a multilevel structure in
which random effects are assumed to be drawn from a common distribution, partially
pooling the information across topics \citep{gelman2007data}.

Generalizing RHEM to contextual RHEM can be done by extending the specification of the relative intensity $\lambda_1(t,\ell,I,J,\beta)=\exp\left[\beta^T\cdot x(t,\ell,I,J)\right]$. In these extended models, we denote the effects formalized in the intensity defined in (\ref{eq:intensity}) as the \emph{global effects}, or \emph{pooled effects}, since they represent aggregate effects over all analyzed papers, irrespective of their context.

\paragraph{Topic-specific intercepts: latent topic-level popularity.}
For each possible label $\ell\in\mathcal{L}$, representing a scientific topic, we introduce a latent popularity parameter $\phi_{\ell}$. A possible reference list $J\subseteq\mathcal{J}$ is associated with a vector $L(J)$ of dimension $|\mathcal{L}|$, whose $\ell$'th component $L(J)_{\ell}$ is the fraction of papers in $J$ that have the label $\ell$, that is
\[
L(J)_{\ell}=\frac{\sum_{j\in J}{\bf 1}\{\textit{label}(j)=\ell\}}{|J|}\enspace.
\]
The term ${\bf 1}\{\textit{condition}\}$ is the indicator function that is one if the \emph{condition} in the argument is true and zero otherwise. To estimate the latent topic popularity, we add the term
\[
\sum_{\ell'\in\mathcal{L}}\phi_{\ell'}\cdot L(J)_{\ell'}
\]
to the argument of the exponential function defining $\lambda_1$ in (\ref{eq:intensity}). If the latent popularity parameter $\phi_{\ell'}$ is positive, then the model suggests that papers in topic $\ell'$ have a higher probability of being cited than papers in the baseline topic (see below). A negative parameter suggests a lower probability. 

\paragraph{Topic-specific slopes: contextual variation of effects across topics.}
Let $x'(t,\ell,I,J)\in\mathbb{R}^{p'}$ be a vector of hyperedge covariates for which we want to assess variation across topics (the entries of $x'$ are typically a subset of the entries of $x$) and let $\beta'\in\mathbb{R}^{|\mathcal{L}|\times p'}$ be a matrix of parameters, whose rows are indexed by topics and whose columns are indexed by the covariates in $x'$, that is, by the effects for which we want to test topic-level variation. The entry in the $q$-th column and the $\ell$-th row of $\beta'$ is the parameter that indicates whether the $q$-th effect is stronger (if positive) or weaker (if negative) for cited papers in topic $\ell$. Since papers in the given list of references $J$ may belong to different topics, we specify the variation dependent on the cited papers by a weighted average, similar to specifying the latent topic-level popularity above. More precisely, to assess the variation of effects across topics, we add the term 
\[
\sum_{\ell'\in\mathcal{L}}{\beta'_{\ell'}}^T\cdot x'(t,\ell,I,J)\cdot L(J)_{\ell'}
\]
to the argument of the exponential function defining $\lambda_1$ in (\ref{eq:intensity}).

\paragraph{Complete model.}
Let $x(t,\ell,I,J)\in\mathbb{R}^p$ be a vector of hyperedge covariates (operationalizing the network effects) and let $x'(t,\ell,I,J)\in\mathbb{R}^{p'}$ be the vector of those covariates for which we want to assess topic-level variation. The complete model is defined by specifying the log relative intensity by:
  \begin{eqnarray*}
    \log\lambda_1(t,\ell,I,J,\theta)&=& \beta^T\cdot x(t,\ell,I,J)\\
    &&+ \sum_{\ell'\in\mathcal{L}}\phi_{\ell'}\cdot L(J)_{\ell'}\\
    &&+ \sum_{\ell'\in\mathcal{L}}{\beta'_{\ell'}}^T\cdot x'(t,\ell,I,J)\cdot L(J)_{\ell'}\\
  \end{eqnarray*}
    where $\theta=(\beta,\phi,\beta')$ are the model parameters. The vector $\beta\in\mathbb{R}^p$ represents the global effects, or pooled effects, the matrix $\beta'\in\mathbb{R}^{|\mathcal{L}|\times p'}$ contains the topic-by-effect interactions (topic-specific slopes for each effect for which we want to estimate contextual variation), and the vector $\phi\in\mathbb{R}^{|\mathcal{L}|}$ contains the topic-level popularity parameters (topic-specific intercepts). The relative intensity $\lambda_1$ is plugged into the partial likelihood (\ref{eq:likelihood}), with the only difference that its parameter vector is now denoted as $\theta$ (and contains the previous $\beta$ as the subset of parameters for the global effects). Parameters are estimated by maximizing the partial likelihood, as discussed before. Estimating all topic-specific intercepts or all topic-specific slopes in one model would create linear dependencies among the explanatory variables. We therefore leave out one reference, or baseline topic. This is similar to estimating regression models having a factor-level variable among the predictors, which requires to leave out the effect of a base level. The choice of the reference topic is essentially arbitrary and actually leads to equivalent models. Since it is convenient to use a ``typical'' topic as the reference, we choose the largest topic in our data (see below when we describe the data).

\subsection{List of network effects}
\label{sec:network_effects}

We use the following network effects $x(t,\ell,I,J)$, which are formally defined in \citet{lerner2025relational,espinosa2025socio,fabbrucci2025co}. We distinguish between the ``core effects'', that is, those effects for which we want to assess variation across topics. and ``control effects'', that is, effects that also---sometimes trivially or obviously---might explain the selection of authors and/or references and that might lead to spurious estimation of core effects, if ignored in the model specification. For convenience, we repeat the formulas defining the core effects in Appendix~\ref{sec:hyperedge_covariates}.

\paragraph{Self-citation.} Following past work analyzing citation networks, we consider self-citation \citep{wolfgang_bibliometric_2004}, a practice that has been linked to increased visibility and downstream citations \citep{fowler_does_2007}, gender differences \citep{king_men_2017, azoulay_self-citation_2020}, and disciplinary variation \citep{snyder_patterns_1998}. Here, incorporating self-citation allows us, for instance, to examine whether authors exhibit higher levels of self-citation in male-dominated topics. Previous work also suggests that larger contexts are characterized by stronger elite control and closure \citep{chu_slowed_2021, linzhuo_social_2020, wittek_development_2023}. Hence, we expect larger topics to exhibit higher levels of self-citation once opportunity structures are accounted for.

Formally, we measure \emph{self-citation} as the ratio of author-reference pairs $(i,j)\in I\times J$ such that $i$ is among the authors of $j$. A positive parameter suggests that authors tend to cite their own prior papers with a higher probability.

\paragraph{Preferential attachment.} It has also long been emphasized that well-connected scholars tend to accumulate further ties, a process later formalized as preferential attachment \citep{merton_matthew_1968, barabasi_emergence_1999, newman_structure_2001}. We examine this dynamic in both collaboration networks, where scientists may preferentially co-author with highly productive peers \citep{newman_structure_2001}, and citation networks, where already well-cited work attracts further citations \citep{eom_characterizing_2011}. This mechanism has far-reaching implications for how scientific contexts evolve: larger and more unequal contexts exhibit lower conceptual diversity \citep{linzhuo_social_2020}, hub-and-spoke collaboration structures are associated with hierarchy and incrementalism \citep{latour_science_1987, latour_laboratory_1986}, and the loss of highly central superstar scientists can increase topical diversity and entry of new scholars \citep{azoulay_superstar_2010, azoulay_does_2019}. Jointly modeling preferential attachment in collaboration and citation networks, therefore, enables a novel, combined perspective on these well-known arguments.

In our models, \emph{preferential attachment (citation)} is a function of the number of prior citations to papers in $J$. A positive parameter suggests that (sets of) papers with a higher number of prior citations are (co-)cited with a higher probability in the future, thereby leading to skewed distributions in the number of citations to papers. The precise specification uses the definition of \emph{geometrically-weighted subset repetition} as formally defined in \citet{fabbrucci2025co}. For a given candidate reference list $J$ at time $t$ we look at all prior publications $j_m=(t_m,\ell_m,I_m,J_m)$ with $t_m<t$, whose reference list $J_m$ overlaps with $J$. Each of these papers is counted with a weight that is increasing in the size of the overlap $|J_m\cap J|$ (see the precise definition given in \citet{fabbrucci2025co} and in Appendix~\ref{sec:hyperedge_covariates}). The geometric weighting is inspired by geometrically weighted terms used in the specification of exponential random graph models \citep{hunter2006inference}.

\emph{Preferential attachment (coauthoring)} is a function of the number of prior publications of authors in $I$. A positive parameter suggests that (sets of) authors with a higher number of prior publications publish or co-publish at a higher rate in the future, thereby leading to skewed distributions in the number of authors' publications. The precise definition is almost identical to the one for preferential attachment for citations, with the only difference that this variable considers the overlap of the authors $I$ with the set of authors $I_m$ of prior publications; the size of the overlap is also geometrically weighted.

\paragraph{Homophily.} Research has consistently shown that ties tend to form between similar actors, a pattern commonly referred to as homophily \citep{mcpherson_birds_2001}. This tendency has been documented across a wide range of social dimensions, including ethnic group membership \citep{smith_ethnic_2016, kroneberg_ethnic_2023, lorenz_link_2021}, race \citep{moody_race_2001, goodreau_birds_2009}, gender \citep{laniado_gender_2016}, or socioeconomic status \citep{lenkewitz_social_2022}. In the science of science, prior work has focused on homophily by gender and status \citep{main_gender_2014, potthoff_is_2017, holman_researchers_2019, burris_academic_2004, ma_boundary_2020}. However, longitudinal statistical analyses of homophily in scientific collaboration remain rare due to scalability and convergence constraints \citep{kronegger_collaboration_2012, espinosa2025socio, stark_meeting_2020}. RHEMs address these limitations while retaining the ability to disentangle multiple mechanisms and account for opportunity structures. We illustrate this by estimating gender homophily in the collaboration networks of oncological researchers and extend the analysis to citation networks, where similar-gender citation patterns can likewise emerge.

In RHEMs, we measure \emph{gender homophily (citation)} as similarity in the ratio of female authors of the citing paper and the cited paper. We first compute the absolute difference of the ratio of females among the citing authors $I$ and the ratio of females among the authors of the cited papers in $J$, which is a measure of gender heterophily. This measure is multiplied by minus one to obtain an indicator of gender homophily. A positive parameter suggests that paper $j$ is more likely to cite paper $j'$ if $j$ and $j'$ have a similar ratio of female authors. In particular, a positive parameter suggests that teams of predominantly female authors tend to cite papers of predominantly female authors and males tend to cite papers of male authors.

\emph{Gender homophily (coauthoring)} measures the fraction of pairs of authors $\{i,i'\}\subseteq I$ among the authors in $I$ that have the same gender. A positive parameter suggests that female authors typically coauthor with other females and males coauthor with males; from another angle, that teams of coauthors are more homogeneous with respect to gender than expected by chance alone.

\paragraph{Control effects.}
\emph{Citation popularity of authors} is the average number of prior citations to any paper of the authors of the cited papers in $J$. A positive parameter suggests that papers of authors whose potentially other papers received a larger number of past citations are cited with a higher probability. In contrast to the preferential attachment effect for citations (defined above), this effect takes into account a possible spillover of popularity from popular papers to other papers of the same author. 
\emph{Author cite paper repetition} is the average number of prior citations of authors in $I$ to papers in $J$. A positive parameter suggests a tendency that the same authors repeatedly cite the same papers, and could again lead to a skewed distribution in the number of citations to papers. This effect requires that the repeated citations to given (sets of) papers are made by the same author(s). In contrast, preferential attachment for citations considers repeated citations to given (sets of) papers, irrespective of the authors of the citing papers. 
\emph{Ratio cited female} measures the ratio of females among the cited authors (main effect of gender on citation probabilities). A positive parameter suggests that papers with a higher ratio of female authors are cited with a higher probability.
\emph{Ratio female authors} operationalizes the main effect of gender on publication probabilities. A positive parameter suggests that females publish more than males.
\emph{Team size of cited paper} is the average number of authors of the papers in $J$. A positive parameter suggests that papers with more authors are cited with a higher probability. 
\emph{Cited paper outdegree} measures the average size of the reference lists of the cited papers in $J$. A positive parameter suggests that papers with a longer reference list (such as review articles) are cited with a higher probability.
\emph{Ratio cited in same topic} is the fraction of cited papers in $J$ that have the same topic as the citing paper $j$. A positive parameter suggests a tendency to cite papers within the same topic. 
\emph{Cite paper and its refs} measures the density of prior citations among the papers in $J$. A positive parameter suggests a tendency to cite papers together with (some of) their references. The effect is related to citation-disruptiveness \citep{park2023papers} in the sense that a tendency to also cite some of the references of a cited paper decreases the disruptiveness of that cited paper.  
\emph{Cite paper of coauthor} computes for each author-reference pair $(i,j)\in I\times J$ we count the number of authors $i'$ of the cited paper $j$, such that the citing author $i$ and the cited author $i'$ have coauthored papers before $t$. We then take the average over these counts over all author-reference pair $(i,j)\in I\times J$. A positive parameter suggests a tendency to cite papers published by prior coauthors.

To attenuate skewed distributions, we transform all effects based on prior publications or citations (i.\,e., all effects except those defined by gender and except \emph{ratio cited same topic}) via the transformation $x\mapsto\log(1+x)$ and then center and standardize these to mean zero and standard deviation equal to one. 

\subsection{Explaining contextual variation: meta-analysis and meta-regression}
\label{sec:metareg}

C-RHEMs yield, for each core effect, a set of topic-specific estimates together with their standard errors. We analyze these estimates in a second step, following an established two-stage approach for comparing statistical network models across contexts \citep{an_multilevel_2015, smith_ethnic_2016, wittek_double_2025}. This second stage serves two purposes: first, to quantify how strongly each effect varies across topics, and second, to assess whether this variation is systematically associated with characteristics of the topics.

\paragraph{Quantifying heterogeneity.} For each of the five core effects, we fit an intercept-only random-effects meta-analytic model to the topic-specific estimates using the \texttt{R} package \texttt{metafor} \citep{viechtbauer_conducting_2010}, in which each estimate is weighted by the inverse of its sampling variance plus the estimated between-topic variance $\tau^2$. The between-topic standard deviation $\tau$ is expressed in the same units as the C-RHEM coefficients and can therefore be compared directly with the corresponding global effect. We complement $\tau$ with $I^2$, the share of the observed dispersion that is attributable to genuine between-topic differences rather than to estimation uncertainty, and with Cochran's $Q$, which tests the null hypothesis that all topics share a common effect. For this step, we reconstruct absolute topic-specific effects for all 32 topics---the global effect plus the respective topic-by-effect interaction---with standard errors computed from the full covariance matrix of the C-RHEM. Absolute effects have the advantage that the heterogeneity statistics do not depend on the choice of reference topic.

\paragraph{Explaining heterogeneity.} To assess whether contextual characteristics account for this variation, we fit a separate random-effects meta-regression for each topic-specific outcome---baseline popularity and each of the five core effects---using the \texttt{R} package \texttt{mvmeta} \citep{gasparrini_multivariate_2012}. In each model, the topic-specific estimates are regressed on three moderators entered simultaneously: topic age, topic size, and the share of male authors (see Section~\ref{sec:data}), each z-standardized so that their coefficients are comparable. Each estimate is thereby decomposed into a systematic component explained by the moderators, a residual between-topic deviation with variance $\tau^2$, and sampling error whose variance is fixed at the squared standard error obtained from the C-RHEM. Parameters are estimated by restricted maximum likelihood (REML), and each topic is weighted by the inverse of its sampling variance plus $\tau^2$. Because the between-topic variance is large relative to the sampling variances (Section~\ref{sec:heterogeneity}), these weights differ only moderately across topics, so that the estimates are not driven by the largest topics.

Because each outcome is modeled separately, the specification involves no between-topic covariance matrix. Effects estimated within the same C-RHEM---for instance, preferential attachment in collaboration and in citation---are correlated, and a multivariate meta-regression could exploit this correlation to gain efficiency. Separate univariate models nonetheless yield valid estimates for each outcome \citep{riley_multivariate_2009}, and they avoid the additional assumptions required to estimate between-topic covariances from only 31 topic-specific estimates. The outcomes of these meta-regressions are the topic-specific deviations from the reference topic, as estimated by the C-RHEMs, so that each meta-regression is based on the 31 non-reference topics; since these differ from the absolute effects only by the global effect, the moderator coefficients have the same substantive interpretation. Finally, comparing the between-topic standard deviation of the intercept-only model with the residual standard deviation after conditioning on the moderators indicates how much of the contextual variation the moderators account for.

\section{Data}
\label{sec:data}

We leverage the PubMed Knowledge Graph (PKG), a comprehensive database containing information on more than 36 million publications in biomedical research \citep{xu_pubmed_2025}. 
We combine collaboration and citation networks derived from the PKG with publication-level topic classifications from OpenAlex (OA), another large-scale bibliometric data source \citep{priem_openalex_2022}. 
OA's topic classification is an automatically generated, hierarchical taxonomy based on publication abstracts, references, and journal outlets \footnote{For details, see: \url{https://help.openalex.org/hc/en-us/articles/24736129405719-Topics}. }. 
It consists of four levels that differ in granularity: (1) 4 Domains, (2) 26 Fields, (3) 252 Subfields, and (4) 4,516 Topics.

We use all OA topics nested within the \textit{oncology} subfield as the contexts across which network mechanisms may vary. 
We chose this analytical sample for two reasons. 
First, oncology is unified by a common focus on the diagnosis and treatment of cancer. Researchers therefore share broad epistemic goals and research infrastructures, reducing differences in network mechanisms that might otherwise arise from fundamentally different disciplinary orientations or organizational principles \citep{becher_significance_1994}.
At the same time, oncology exhibits substantial heterogeneity and occupies a middle ground between research areas requiring few resources and characterized by predominantly single-author publications, such as philosophy, and domains involving very large teams and substantial resources, such as high-energy physics. While some oncology topics are typically pursued by small- to medium-sized teams (e.g., research on the biochemical underpinnings of cancer), others are organized around large laboratories or hospital-based clinical research \citep{maman_history_2018, lazega_organizational_2006}. 
This combination of shared disciplinary foundations and meaningful organizational variation makes oncology a suitable setting for illustrating our methodology, which is designed to identify differences in network mechanisms across contexts.

Second, with about 590,000 research articles, oncology is computationally demanding but still feasible to analyze on a local machine using RHEMs.\footnote{Specifically, RHEM covariates for our analysis were computed with \texttt{eventnet} on a Lenovo ThinkPad P17 Gen 2i with an 11th Gen Intel Core i9 processor and 128 GB RAM, running Ubuntu 24.04.4 LTS in less than two hours. Given the covariates, the estimation of model parameters with the \texttt{coxph} function in the \texttt{R} package \texttt{survival} takes less than 200 seconds on a Lenovo ThinkPad X1 G13 with an Intel Core Ultra 7 258V processor and 32 GB RAM, running Windows 11.} Larger datasets would typically require access to a high-performance computing cluster. We therefore use oncology to demonstrate a workflow that researchers can readily reproduce without specialized computing infrastructure. Moreover, many collaboration and citation network datasets used in research on scientific and cultural production are of comparable or smaller size \citep{cheng_how_2023, heiberger_facets_2021, higham_patent_2021, burgdorf_communities_2024}, facilitating both substantive comparisons and the direct application of the code provided here.

We applied three filters to the full PKG before restricting it to oncology. First, we retained only publications from 1980 to 2023, because data quality is substantially lower in earlier years, with high levels of missing information for key variables such as author names. Second, we kept only publications classified as ``research articles'' in the PKG, excluding book chapters, conference proceedings, and other publication types. Third, for each publication, we calculated the share of references pointing to other articles in the PKG (i.e., the inward citation rate) and averaged it for each OA topic across the full PKG. Because inward citation reflects how comprehensively a topic is represented in PubMed, we excluded topics with an average rate below 80\%, as their citation networks are potentially incomplete. Restricting the filtered data to the 34 OA topics in the oncology subfield yields 590,547 research articles. Among these topics, only ``Platinum-Based Cancer Chemotherapy'' (T10163) falls below the inward-citation threshold (approximately 57\%, five research articles), and we excluded it. Finally, we excluded topics with fewer than 1,000 publications, thereby removing ``Nanotechnology in Cancer Treatment and Diagnosis'' (T13434), which had 630 publications. The resulting dataset comprises 589,912 publications in 32 topics, written by 1,455,842 distinct authors and connected by 5,236,288 citation links within the dataset.

To respect this network boundary, we retain, for each publication $j=(t,\ell,I,J)$, only those references in $J$ that are in our dataset. Publications whose reference list thereby becomes empty are not modeled as events but remain in the data as possible references of other papers, which leaves 466,135 publication events. Case-control sampling (see below) adds five non-events per event, yielding 2,796,810 observations. Of these, 227,249 observations lack gender information for at least one covariate and are excluded from estimation, leaving 429,691 events and 2,569,561 observations (Table~\ref{table:coef_population-level}).

Table \ref{tab:fields} shows how topics embedded in oncology vary in their characteristics. The table is ordered by topic size, which ranges from ca.\ 43k (``Pancreatic Cancer Research and Treatment'', which we select as the reference topic) to ca.\ 2.3k (``Role of Clusterin in Cancer and Disease'') research articles; we consider topic size as the first contextual moderator. The share of inward citations is consistently high, with a minimum of 84\% and a mean of 94.8\%. Topics vary in the share of male author names detected (male names are divided by the sum of male and female names), which is the second contextual moderator, and in the median publication year within each topic. Note that we considered topic age as the third contextual moderator in meta-regressions, computed by subtracting the median publication year from 2022. As in many previous studies, we inferred gender from first names \citep[e.g.,][]{hofstra_diversityinnovation_2020}. We used the Wikidata-based classifier developed by \citet{berube_wiki-gendersort_2020}, which outperforms alternatives such as Genderize.io in benchmark evaluations \citep{sebo_performance_2021}. Because the classifier is free, open-source, and based on publicly available data, it ensures a transparent and reproducible classification procedure. It assigns a binary gender to 79.5\% of the 3.8 million author-name instances in our data (53.1\% male and 26.4\% female); the remaining names are unisex (4.5\%), consist of initials only (0.8\%), are unknown to the classifier (8.4\%), or could not be matched (6.8\%). 

{\setlength{\tabcolsep}{3pt}
\begin{longtable}{p{4.5cm}rrrrr}
\caption{Summary of research topics}
\label{tab:fields} \\
\toprule
\textbf{Topic} & \textbf{OA Topic ID} & \textbf{No. Papers} & \textbf{Inward Share} & \textbf{Median Year} & \textbf{Male (\%)} \\
\midrule
\endfirsthead
\multicolumn{6}{c}{\small\tablename\ \thetable{} -- continued} \\[2pt]
\toprule
\textbf{Topic} & \textbf{OA Topic ID} & \textbf{No. Papers} & \textbf{Inward Share} & \textbf{Median Year} & \textbf{Male (\%)} \\
\midrule
\endhead
\midrule
\multicolumn{6}{r}{\small\textit{Continued on next page}} \\
\endfoot
\bottomrule
\endlastfoot

{\footnotesize Pancreatic Cancer Research and Treatment} & 10231 & 43{,}201 & 0.97 & 2015 & 77 \\[3pt]
{\footnotesize Mechanisms of Multidrug Resistance in Cancer} & 10570 & 42{,}104 & 0.95 & 2005 & 69 \\[3pt]
{\footnotesize The p53 Signaling Network in Cancer Research} & 10583 & 33{,}402 & 0.97 & 2004 & 67 \\[3pt]
{\footnotesize Colorectal Cancer Research and Treatment} & 10335 & 29{,}727 & 0.96 & 2014 & 79 \\[3pt]
{\footnotesize Viral-Related Cancers in Immunocompromised Patients} & 10685 & 27{,}284 & 0.96 & 2007 & 66 \\[3pt]
{\footnotesize Cancer Immunotherapy} & 10158 & 24{,}481 & 0.97 & 2020 & 65 \\[3pt]
{\footnotesize Global Cancer Incidence and Mortality Patterns} & 10556 & 24{,}294 & 0.91 & 2011 & 50 \\[3pt]
{\footnotesize Cancer Stem Cells and Tumor Metastasis} & 10336 & 23{,}971 & 0.98 & 2015 & 63 \\[3pt]
{\footnotesize Global Trends in Colorectal Cancer Research} & 10552 & 23{,}706 & 0.94 & 2013 & 67 \\[3pt]
{\footnotesize Bisphosphonates in Cancer Treatment and Bone Health} & 11221 & 23{,}624 & 0.95 & 2007 & 70 \\[3pt]
{\footnotesize Melanoma} & 10392 & 23{,}152 & 0.92 & 2011 & 65 \\[3pt]
{\footnotesize Role of Fibroblast Activation in Cancer Progression} & 12505 & 23{,}083 & 0.93 & 2006 & 69 \\[3pt]
{\footnotesize Chemotherapy-Induced Peripheral Neuropathy in Cancer Treatment} & 11752 & 22{,}266 & 0.90 & 2001 & 70 \\[3pt]
{\footnotesize Cancer Survivorship and Quality of Life} & 10178 & 20{,}701 & 0.90 & 2016 & 44 \\[3pt]
{\footnotesize Cancer of Unknown Primary Site} & 12379 & 19{,}519 & 0.94 & 2002 & 75 \\[3pt]
{\footnotesize Role of STAT3 in Cancer Inflammation and Immunity} & 11503 & 18{,}408 & 0.97 & 2007 & 67 \\[3pt]
{\footnotesize Impact of COVID-19 on Cancer Patients and Care} & 11296 & 18{,}065 & 0.95 & 2021 & 62 \\[3pt]
{\footnotesize Advancements in Colorectal Cancer Research} & 11067 & 15{,}004 & 0.95 & 2012 & 72 \\[3pt]
{\footnotesize Polyomavirus-Associated Carcinogenesis and Clinical Manifestations} & 11827 & 14{,}784 & 0.97 & 2008 & 66 \\[3pt]
{\footnotesize HER2 Signaling in Breast Cancer Treatment} & 10755 & 14{,}748 & 0.96 & 2011 & 63 \\[3pt]
{\footnotesize Chimeric Antigen Receptor T Cell Therapy} & 11491 & 14{,}123 & 0.97 & 2019 & 63 \\[3pt]
{\footnotesize Inflammation's Role in Cancer Development and Progression} & 11355 & 10{,}955 & 0.98 & 2019 & 72 \\[3pt]
{\footnotesize Impact of Obesity on Cancer Risk and Prognosis} & 12103 & 10{,}721 & 0.95 & 2010 & 51 \\[3pt]
{\footnotesize Lymphangiogenesis in Cancer Metastasis and Disease} & 11469 & 10{,}665 & 0.95 & 2014 & 63 \\[3pt]
{\footnotesize Management of Febrile Neutropenia in Cancer Patients} & 12172 & 10{,}506 & 0.95 & 2005 & 66 \\[3pt]
{\footnotesize Role of Chemokine Receptors in Cancer Metastasis} & 11532 & 10{,}118 & 0.97 & 2009 & 66 \\[3pt]
{\footnotesize Vascular Tumors and Angiosarcoma} & 12278 & 9{,}952 & 0.94 & 2009 & 70 \\[3pt]
{\footnotesize Small Cell Lung Cancer} & 12334 & 8{,}692 & 0.95 & 2010 & 74 \\[3pt]
{\footnotesize Poly(ADP-ribose) Polymerase Inhibition in Cancer Therapy} & 11914 & 7{,}880 & 0.96 & 2016 & 61 \\[3pt]
{\footnotesize Tumors of the Ear and Temporal Bone} & 14326 & 6{,}007 & 0.93 & 2006 & 77 \\[3pt]
{\footnotesize Cancer Chemopreventive Properties of Saffron} & 12758 & 2{,}505 & 0.84 & 2017 & 62 \\[3pt]
{\footnotesize Role of Clusterin in Cancer and Disease} & 14364 & 2{,}264 & 0.97 & 2009 & 67 \\[3pt]

\end{longtable}}

\paragraph{Estimation.} Computation of explanatory variables is done with the \texttt{eventnet} software \citep{lerner2023relational}. We perform nested case-control sampling \citep{borgan1995methods,lerner2025relational} with five non-events (that is, alternative, randomly selected lists of authors and papers that could have been the authors, respectively, could have been cited) per publication. Once the explanatory variables have been computed, we estimate model parameters with the \texttt{coxph} function in the \texttt{R} package \texttt{survival} \citep{therneau2000modeling,therneau2024package}.

\paragraph{Use of AI tools.} We used a large language model (Anthropic's Claude) to assist with drafting and revising portions of the manuscript text and with parts of the replication code, including the script computing the heterogeneity statistics reported in Section~\ref{sec:heterogeneity}. All text and code were reviewed, revised, and verified by the authors, who take full responsibility for the content of this article.

\section{Results}

\subsection{Global effects} 

\begin{table}
\begin{center}
\begin{tabular}{l r}
\hline
Preferential attachment (citation) & $0.357 \; (0.002)^{***}$  \\
Preferential attachment (coauthoring) & $0.204 \; (0.002)^{***}$  \\
Gender homophily (citation) & $0.360 \; (0.012)^{***}$  \\
Gender homophily (coauthoring) & $0.262 \; (0.008)^{***}$  \\
Self-citation & $0.058 \; (0.001)^{***}$  \\
Citation popularity of authors & $-0.062 \; (0.004)^{***}$ \\
Author cite paper repetition & $-0.104 \; (0.001)^{***}$ \\
Ratio female authors & $0.085 \; (0.009)^{***}$  \\
Ratio cited female & $0.714 \; (0.011)^{***}$  \\
Team size of cited paper & $0.118 \; (0.002)^{***}$  \\
Cited paper outdegree & $0.038 \; (0.003)^{***}$  \\
Ratio cited in same topic & $4.073 \; (0.007)^{***}$  \\
Cite paper and its refs & $0.002 \; (0.001)^{*\phantom{**}}$    \\
Cite paper of coauthor & $0.005 \; (0.002)^{***}$  \\
\hline
AIC                            & $2{,}876{,}250.470$             \\
Num. events                    & $429{,}691$                  \\
Num. obs.                      & $2{,}569{,}561$                 \\
Missings                       & $227{,}249$                  \\
\hline
\multicolumn{2}{l}{\scriptsize{$^{***}p<0.001$; $^{**}p<0.01$; $^{*}p<0.05$}}
\end{tabular}
\caption{Estimated parameters and standard errors (in parentheses) of RHEM without any topic-level variation. Missing observations are due to missing gender information and comprise 36,444 events and 190,805 sampled non-events.}
\label{table:coef_population-level}
\end{center}
\end{table}

We begin by estimating a model that assumes all network mechanisms are constant across oncological research topics. 
Table~\ref{table:coef_population-level} reports the resulting parameters of the global effects, which serve as a benchmark against which we assess contextual variation below.

Consistent with previous work in the science of science, we find strong evidence for preferential attachment in both citation and collaboration networks. Papers that have accumulated more citations are more likely to be cited in the future, while scientists with more previous publications are more likely to attract additional collaborators. We also find positive gender homophily in both collaboration and citation networks, indicating that researchers tend to collaborate with and cite teams with a similar gender composition. Finally, we observe significant self-citation. Thus, all focal network mechanisms are supported under the assumption of a common generative process across all oncology topics.

Effects have varying sizes. For example, the parameter $\beta=0.357$ for preferential attachment in citations implies that if the respective covariate increases by one standard deviation, then the expected intensity gets multiplied by a factor of $\exp(0.357)=1.43$, that is, an increase by 43\%. For preferential attachment in coauthoring, the factor is $\exp(0.204)=1.23$, and for self-citation, the factor is $\exp(0.058)=1.06$, corresponding to an increase of 6\% in the expected intensity, implied by one standard deviation increase in the covariate. Note that while some effects of substantive interest are larger than others, this does not imply that they are theoretically irrelevant. For instance, preferential attachment is long known as one of the strongest organizing principles of networks \citep{newman_scientific_2001, barabasi_emergence_1999}, but self-citations are also a meaningful phenomenon investigated by a range of studies because they point to the hoarding of citations by elites even though they are probably less consequential for the overall structure of the network \citep{azoulay_self-citation_2020, fowler_does_2007}.  
Some of the control variables have very large effect sizes. For example, the parameter of \emph{ratio cited same topic} is 4.073, implying a multiplication of the intensity by a factor of $\exp(4.073)=58.73$ when the covariate changes from zero to one. (Recall that \emph{ratio cited same topic}, as well as the gender-based covariates, are not standardized since they have a ``natural'', or interpretable scale. In the case of \emph{ratio cited same topic} the switch from zero to one corresponds to the hypothetical change from a reference list in which no referenced paper has the citing paper's topic to a reference list that includes only papers in the citing paper's topic.) The finding that papers have a preference to cite other papers with the same topic can be seen as highly expected -- however, the strength of the effect underlines the need to control for such, sometimes obvious or trivial, regularities in a model explaining coauthoring and citation.

The global model provides an important baseline. However, it implicitly assumes that collaboration and citation evolve according to identical network mechanisms across all research topics. 
Network ecology proposes that this assumption is unlikely to hold. 
We therefore next apply the extended RHEM framework discussed above by allowing network mechanisms to vary across scientific topics.
As discussed above, we select Topic 10231 (\emph{Pancreatic Cancer Research and Treatment}), which constitutes the largest and one of the most representative research areas in our sample, as the reference category.

\subsection{Model fit improvement by C-RHEM over the global model}
\label{sec:model_fit}

We estimate four different C-RHEMs: one in which only paper popularity varies across topics (topic-specific intercepts but no topic-specific slopes), and three extensions of this model in which, in addition, one of the three core mechanisms---preferential attachment, gender homophily, or self-citation---varies across topics. Table~\ref{table:AIC_improvements} shows that each C-RHEM improves the AIC over the global model reported in Table~\ref{table:coef_population-level}, and that adding topic-specific slopes for each core mechanism yields a further improvement over the model with topic-specific popularity only. Varying preferential attachment brings the largest additional improvement, followed by varying self-citation and varying gender homophily. The full parameter tables of these models are reported in Online Resource~1, and the topic-specific estimates are displayed in Appendix~\ref{sec:topic_estimates}.

\begin{table}
\begin{center}
\begin{tabular}{l r}
\hline
C-RHEM variant & $\Delta$ AIC\\
\hline
topic-level popularity (fixed intercepts) & 8,639\\
\hline
topic-level popularity and varying preferential attachment & 17,095\\
topic-level popularity and varying gender homophily & 9,314\\
topic-level popularity and varying self-citation & 9,763\\
\hline
\end{tabular}
\caption{Improvements in AIC by various C-RHEM over the baseline model without any topic-level variation, reported in Table~\ref{table:coef_population-level}. The baseline model has an AIC equal to $2{,}876{,}250$. Note that the first model (``topic-level popularity'') is nested in each of the three others.}
\label{table:AIC_improvements}
\end{center}
\end{table}

\subsection{How strongly do network effects vary across topics?}
\label{sec:heterogeneity}

Table~\ref{table:tau} summarizes the between-topic heterogeneity of the five core effects, obtained with the procedure described in Section~\ref{sec:metareg}. 
Three results stand out. First, the variation across topics is not an artifact of estimation uncertainty: $I^2$ exceeds $90$\% for all five effects, and $Q$ rejects the null of homogeneous effects in every case. Second, the magnitude of the contextual variation is comparable to that of the average effect itself. The between-topic standard deviation ranges from roughly $30$\% of the corresponding global estimate (preferential attachment in coauthoring) to around $70$\% (gender homophily in coauthoring). For gender homophily in coauthoring, for instance, $\tau=0.183$ implies that a topic one standard deviation above the mean exhibits a homophily intensity approximately $1.44$ times that of a topic one standard deviation below it. Third, and in contrast, the absolute magnitude of the variation in self-citation remains modest: although $\tau=0.025$ is large relative to the small global estimate of $0.058$, the corresponding difference between topics one standard deviation apart amounts to a factor of only about $1.05$.

\begin{table}[h!]
\begin{center}
\begin{tabular}{l r r r r r}
\hline
Network effect & Global $\beta$ & $\tau$ & Residual $\tau$ & $I^2$ (\%) & $Q$ ($df=31$) \\
\hline
Preferential attachment (citation) & $0.357$ & $0.126$ & $0.099$ & $99.4$ & $5394.3^{***}$ \\
Preferential attachment (coauthoring) & $0.204$ & $0.060$ & $0.035$ & $98.5$ & $1727.4^{***}$ \\
Gender homophily (citation) & $0.360$ & $0.202$ & $0.137$ & $90.7$ & $331.1^{***}$ \\
Gender homophily (coauthoring) & $0.262$ & $0.183$ & $0.153$ & $95.8$ & $464.4^{***}$ \\
Self-citation & $0.058$ & $0.025$ & $0.016$ & $97.1$ & $1031.5^{***}$ \\
\hline
\multicolumn{6}{l}{\scriptsize{$^{***}p<0.001$}}
\end{tabular}
\caption{Between-topic heterogeneity of the five core network effects. $\tau$ is the estimated between-topic standard deviation from an intercept-only meta-analytic model fitted to the 32 topic-specific estimates; residual $\tau$ is the corresponding quantity after conditioning on topic age, topic size, and the share of male authors. $I^2$ gives the proportion of total variation attributable to between-topic differences; $Q$ tests the null of homogeneous effects across topics. Global $\beta$ is reproduced from Table~\ref{table:coef_population-level} for comparison.}
\label{table:tau}
\end{center}
\end{table}

The topic-specific estimates underlying these statistics are displayed in Appendix~\ref{sec:topic_estimates}. They also show that baseline citation popularity varies considerably across topics---most visibly for research on the impact of COVID-19 on cancer care, which attracts substantially more citations than the reference topic, likely reflecting an exogenous surge in attention during the pandemic. Because the C-RHEMs control for such differences through topic-specific intercepts, the heterogeneity in Table~\ref{table:tau} reflects differences in network mechanisms rather than in the overall attractiveness of research topics.

\subsection{Explaining contextual variation}
\label{sec:moderators}

\begin{figure}[h!]
    \centering
    \includegraphics[width=0.75\linewidth]{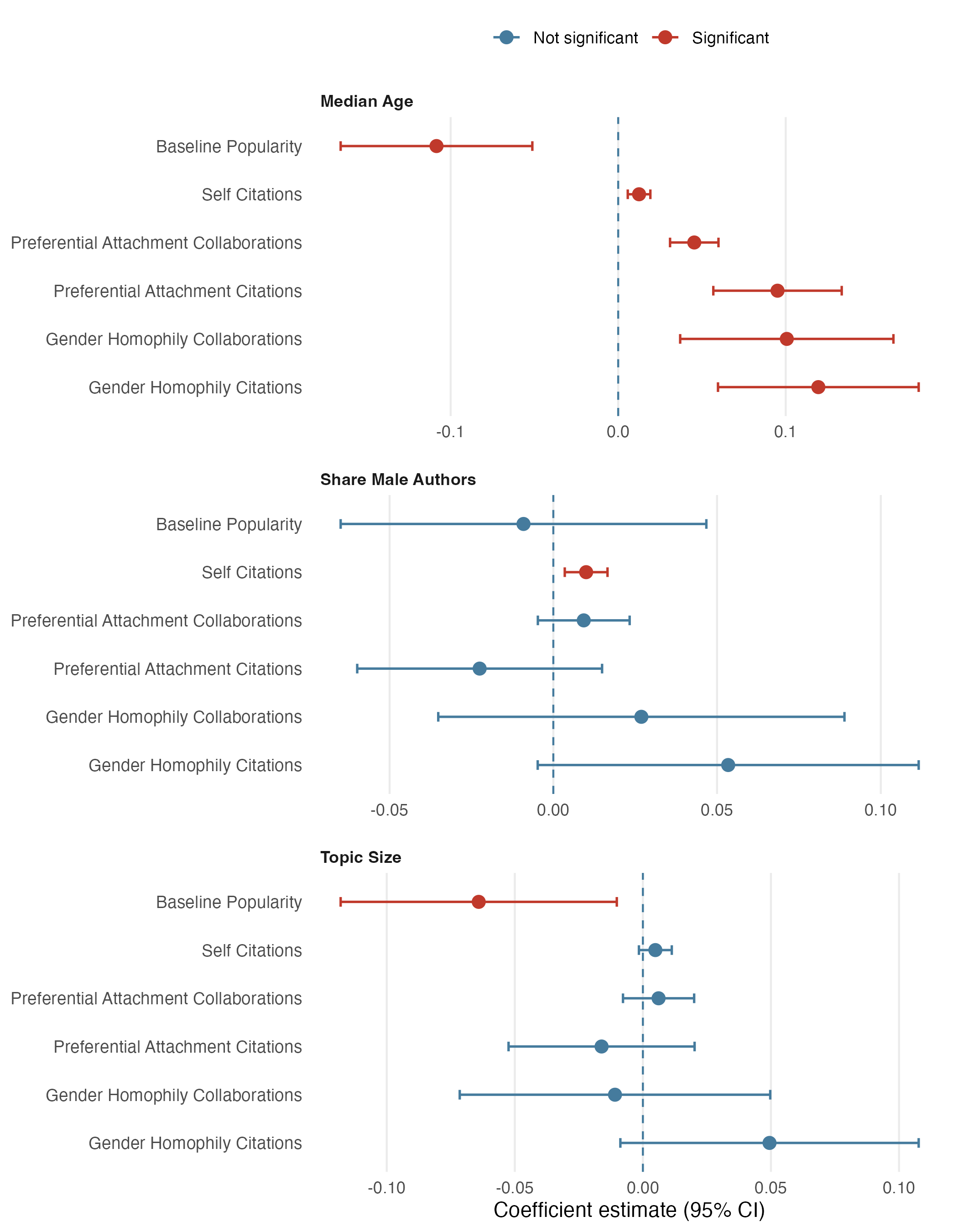}
    \caption{\textbf{Meta-regressions of topic-specific network effects on contextual moderators.} Each panel depicts how one contextual moderator is related to each network effect, controlling for the other two moderators. All moderators are z-standardized to facilitate comparison of coefficient sizes. Estimates are obtained by restricted maximum likelihood with the \texttt{R} package \texttt{mvmeta}; see Section~\ref{sec:metareg} and \citet{gasparrini_multivariate_2012} for details.}
    \label{fig:regressions}
\end{figure}

Having established that network effects vary substantially across topics, we turn to the question of whether this variation can be explained by characteristics of the topics. Figure~\ref{fig:regressions} reports meta-regressions of the topic-specific effects on topic age, the share of male authors, and topic size (Section~\ref{sec:metareg}).

Most notably, topic age is positively associated with preferential attachment in both collaboration and citation networks, as well as with self-citation. Moreover, gender homophily in citations and collaborations is stronger in older topics. This result indicates that authors in older topics prefer to collaborate with others who have the same gender and tend to cite contributions by teams with a similar gender composition more often. Older research topics, therefore, tend to exhibit more closure according to gender and stronger cumulative advantage. The contrast between the long-established topic \emph{Cancer of Unknown Primary Site} and the recently emerging COVID-19-related research illustrates this pattern: the former displays substantially stronger preferential attachment, self-citation, and gender homophily than the latter (see Figures~\ref{fig:base_and_self}--\ref{fig:prefat_fields} in Appendix~\ref{sec:topic_estimates}).

Demographic composition also appears to matter, though the evidence is stronger for some mechanisms than others. Topics with a higher proportion of male authors exhibit significantly stronger self-citation. For gender homophily, the association points in the same direction in both collaboration and citation networks, consistent with the bivariate correlations reported in Appendix~\ref{sec:descriptives} ($r \approx 0.32$--$0.43$, Figure~\ref{fig:context_moderators}), but does not reach conventional statistical significance. This lack of statistical significance is likely due to the moderate number of topics in our sample (31 non-reference topics in the meta-regressions). We therefore treat these associations as suggestive evidence for a link between topics' demographic composition and gender homophily; further research should revisit these initial findings using a dataset with a larger number of topics (for instance, the entire PubMed Knowledge Graph).

Contrary to our expectations, topic size exhibits mostly insignificant associations with the network mechanisms considered here. The only statistically significant association is a negative link between baseline popularity and size, suggesting that papers in larger topics receive fewer citations than predicted by the other model terms. However, in line with the notion that larger topics are crystallized communities organized around elites, we find moderate positive correlations between topic size and self-citations as well as preferential attachment in collaborations (see Figure~\ref{fig:context_moderators} in the Appendix). In addition, we find a moderate bivariate relation between gender homophily in citations and topic size.

How much of the contextual variation do these three moderators account for? Comparing the unconditional and residual between-topic standard deviations in Table~\ref{table:tau}, the reduction in between-topic variance is substantial across all five effects, ranging from roughly a third for gender homophily in coauthoring to around two thirds for preferential attachment in coauthoring. Topic age, size, and gender composition thus capture a considerable share of the contextual variation in every effect we examine, while leaving a non-negligible remainder unexplained.

Taken together, these analyses demonstrate that the heterogeneity uncovered by contextual RHEMs is not merely random variation across scientific topics. Instead, a substantial share can be explained by theoretically meaningful contextual characteristics. More broadly, the results illustrate how C-RHEMs can operationalize the central ideas of network ecology by estimating and explaining how network-generating mechanisms vary across large-scale relational environments.

\section{Discussion}

Research in scientometrics and the science of science has documented the mechanisms that generate collaboration and citation networks in considerable detail, including cumulative advantage \citep{merton_matthew_1968, price_general_1976}, gender homophily \citep{main_gender_2014, potthoff_is_2017}, and self-citation \citep{wolfgang_bibliometric_2004, azoulay_self-citation_2020}. It has also established that research areas differ in their levels of citation, self-citation, and collaboration \citep{waltman_review_2016, aksnes_macro_2003, kronegger_collaboration_2012}. Whether the \emph{mechanisms} themselves differ across research areas has received far less attention, mainly because statistical network models that estimate several mechanisms jointly are hard to apply to large bibliometric data and even harder to compare across subnetworks. This article addresses that gap. We introduced contextual relational hyperevent models (C-RHEMs), which retain the key advantage of RHEMs---modeling each publication as a single hyperevent that jointly links its authors to its references, rather than decomposing it into dyadic ties---and which estimate topic-specific intercepts and slopes for selected network effects within a single partial-likelihood model. We combined them with a meta-analytic second stage that quantifies the resulting heterogeneity and relates it to characteristics of research topics.

Applied to about 590,000 oncology publications in 32 OpenAlex topics, the approach yields three main results. First, all three mechanisms vary substantially across topics. Between-topic standard deviations amount to 30 to 70 percent of the corresponding pooled effects, and with $I^2$ above 90 percent for every effect, almost all of the observed dispersion reflects differences between topics rather than estimation uncertainty. This is notable because the topics all belong to one subfield with a shared scientific objective; the variation across more distant research areas is likely larger still. Second, topic age is the most consistent correlate of this variation. Older topics show stronger preferential attachment in both collaboration and citation, stronger self-citation, and stronger gender homophily. Topics with a higher share of male authors show stronger self-citation and, more tentatively, stronger gender homophily. Third, once age and gender composition are taken into account, topic size shows few systematic associations, although its bivariate correlations with self-citation, preferential attachment in collaboration, and citation gender homophily point in the expected direction. Together, the three characteristics account for between a third and two-thirds of the between-topic variance, leaving a non-negligible remainder for other factors.

The prominence of topic age is consistent with Crane's \citeyearpar{crane_invisible_1972} account of invisible colleges that mature from small, open groups into consolidated communities in which recognition and resources are increasingly concentrated. It also complements work documenting increasing inequality and slower turnover of ideas in large scientific fields \citep{chu_slowed_2021, linzhuo_social_2020}. Whereas that work emphasizes field size, our results suggest that, at least within a subfield, age is the more consistent correlate of cumulative advantage and closure. Similarly, the finding that gender homophily is stronger in older topics fits evidence that gender homophily in citation is weaker among more recent cohorts of scientists \citep{zhou_gender_2024}. Because older topics are populated by earlier cohorts, separating topic age from cohort composition is an important task for future research.

The results have several implications for the study and use of bibliometric data. The first concerns the interpretation of citation indicators. Field normalization adjusts for differences in the \emph{level} of citation activity across research areas \citep{waltman_review_2016}, and the topic-specific intercepts of C-RHEMs play a comparable role within the model. Our results indicate that the \emph{processes} through which citations accumulate differ as well. How strongly prior citations attract further citations, and how strongly authors cite their own work, vary considerably across topics even within a single subfield. Two papers with the same normalized citation score may therefore have reached it through different mixtures of cumulative advantage, self-citation, and other mechanisms, depending on where they are located. Normalization equalizes levels but not these mixtures, which is relevant whenever citation indicators are used to compare research areas, institutions with different topical profiles, or individuals; that is, in most bibliometric metrics for academic success.

Second, network mechanisms estimated on pooled data are averages over processes that can differ markedly across the parts of a network. A global preferential attachment or homophily parameter is a useful summary, but our results show that it can mask between-topic standard deviations of more than half its own size. Conclusions about cumulative advantage, gender homophily, or self-citation drawn from one research area, or from data pooled across many, need not carry over to others. This applies to the large literature on gender differences in collaboration and citation in particular \citep{kim_gendered_2022, zhou_gender_2024, santos_homophily_2024}: the extent of gender homophily is itself a property of the research area, and it is associated with the area's gender composition and age. Studies that compare gender gaps across fields or over time may therefore benefit from modeling the context-dependence of homophily explicitly rather than treating it as a constant.

Third, the results bear on the choice of the unit of analysis. Much scientometric work uses journal-based subject categories or broad disciplines as contexts. C-RHEMs make it possible to work with fine-grained, overlapping, publication-level classifications such as the OpenAlex topics used here \citep{priem_openalex_2022}, because topic membership enters the model as the share of a paper's references in each topic. The finding of large heterogeneity at this fine-grained level suggests that classification granularity matters for network analysis and not only for indicator normalization, and C-RHEMs provide a way to assess this systematically by re-estimating the same model under alternative classifications.

Compared with existing approaches, C-RHEMs have three practical advantages. First, unlike studies that estimate separate stochastic actor-oriented models for each discipline \citep{kronegger_collaboration_2012, ferligoj_scientific_2015}, they estimate all topic-specific effects within one model, so the estimates are on a common scale and can be compared and analyzed with standard meta-analytic tools. Second, unlike configuration models and related null models that scale to large bibliometric data \citep{newman2001random, uzzi2013atypical}, they estimate several mechanisms jointly, control for one another, and allow each mechanism's strength to vary. Third, they represent each publication as a single hyperevent that links its authors to its references \citep{lerner2023relational, lerner2025relational}, which avoids the information loss and structural artifacts of projecting publications onto dyadic ties \citep{battiston2020networks, lerner2023micro} and allows collaboration and citation to be modeled together. With covariate computation taking less than two hours and estimation a few minutes on standard hardware, the workflow can be reproduced without access to a computing cluster. Bibliometric datasets of comparable or smaller size are common in the literature \citep{heiberger_facets_2021, higham_patent_2021, burgdorf_communities_2024}, so the approach can be applied directly to other fields, other classifications, or other contextual units such as countries, institutions, or journals, using \texttt{eventnet} to compute the covariates and our \texttt{R} scripts for estimation and meta-analysis.

Within science, the idea of moderating field contexts has a long qualitative tradition. Invisible colleges \citep{crane_invisible_1972}, epistemic cultures \citep{cetina_epistemic_1999}, paradigm communities \citep{kuhn_structure_1970}, and laboratory life \citep{latour_laboratory_1986} all describe research areas as social worlds with their own ways of collaborating, attributing credit, and building on prior work. Our results can be read as a first quantitative step toward testing such accounts at scale: if research areas differ in their epistemic and social organization, these differences should leave traces in the mechanisms that generate their collaboration and citation networks. C-RHEMs offer a way to measure these traces and to relate them to observable characteristics of research areas, and could eventually be linked to more direct measures of epistemic culture, such as resource intensity, team structure, or degree of consensus.

Although we developed C-RHEMs for bibliometric data, the underlying question---whether, and why, the mechanisms that generate social networks differ across the contexts in which they unfold---extends well beyond scientometrics. Network ecology \citep{mcfarland_network_2014, doehne_network_2024} has posed this question for adolescent friendship networks across schools, and studies of organizations and communities have long suggested that relational processes respond to the size, age, and composition of their settings. As a recent handbook puts it, researchers must understand how relationships operate \emph{within particular subgroups} of a population before they can ``properly appreciate the significance of network data'' \citep[][p.~9]{crossley_introduction_2025}. Such a view resonates with sociological accounts that emphasize the mutual constitution of local cultures and social relations \citep{bourdieu_field_1993, fine_hinge_2021}.

The same reasoning applies to other domains of cultural production. Patents, films, music, books, and video games all generate multi-mode networks linking creators to products and products to their predecessors \citep{higham_patent_2021, bottero_worlds_2011, wittek_networks_2025, burgdorf_communities_2024}, and genres, markets, or scenes provide natural contexts across which mechanisms such as cumulative advantage or homophily may vary. Comparing network mechanisms across such domains could eventually support a more general understanding of how the age, size, and composition of creative fields shape the social organization of creative work.

Our study has several limitations. First, it examines a single subfield. This was a deliberate choice for a tractable illustration with shared epistemic foundations, but the magnitude of heterogeneity and the role of individual moderators may differ elsewhere. Second, the meta-regressions are cross-sectional across 31 topics. The associations with topic age, size, and gender composition should therefore be read as descriptive rather than causal, and the small number of contexts limits statistical power, as the non-significant but consistent associations between gender composition and homophily illustrate. Applying C-RHEMs to a larger number of topics, for instance the entire PubMed Knowledge Graph, and modeling how mechanisms change as topics age would allow stronger tests. Third, our contexts are topics assigned by the OpenAlex classification algorithm rather than self-identified research communities. C-RHEMs accommodate overlap between topics, but the estimates are conditional on this classification; defining contexts through community detection in the collaboration or citation network, or comparing results across classification systems, are natural extensions. Fourth, we restricted reference lists to publications within the dataset. Citations to work outside oncology, or outside PubMed, are therefore not modeled, which may affect topics that draw heavily on neighboring fields. Fifth, our reliance on name-based gender inference carries limitations common to this approach. Such methods infer rather than observe gender and assign individuals to a binary category that may not reflect self-identified gender. Classification accuracy varies across naming traditions and tends to be lower for non-Western and non-Anglophone names, which can introduce non-random measurement error \citep{lockhart_name-based_2023}. These issues may interact with the substantial share of observations excluded because of missing gender information (Table~\ref{table:coef_population-level}) in ways we have not fully characterized. Our estimates of gender homophily should therefore be interpreted as approximate patterns contingent on the assumptions of name-based inference.

The approach also involves choices that users should be aware of. We estimated topic-specific parameters as fixed effects, which is appropriate when every context contains enough events; with many small contexts, a multilevel specification with partial pooling would be preferable \citep{gelman2007data}. The two-stage design separates estimation of topic-specific effects from their comparison, propagating first-stage uncertainty through inverse-variance weights. Its power to detect moderators is limited by the number of contexts rather than the number of publications, which is why some associations in our analysis remain suggestive despite more than two million observations.

Future work could examine further contextual characteristics, such as funding regimes, the share of clinical versus basic research, geographic composition, or the degree of interdisciplinarity of a topic. Methodologically, extending C-RHEMs to multiple or hierarchically nested contexts---for instance, topics within subfields within fields---and to random topic-specific effects for settings with many small contexts are promising directions.

In sum, C-RHEMs extend relational hyperevent models to estimate how network mechanisms vary across the parts of large bibliometric networks while preserving the higher-order structure of scientific publishing. Our application shows that preferential attachment, gender homophily, and self-citation differ substantially across research topics, even within a single subfield, and that much of this variation is associated with the age and gender composition of topics. For scientometrics, this suggests treating the mechanisms behind collaboration and citation, and not only their levels, as properties of research areas. C-RHEMs offer a scalable way to do so, and potentially a starting point for a comparative, quantitative study of how contexts shape the social organization of science and other creative fields.

\section*{Statements and Declarations}

\paragraph{Competing interests.} The authors have no competing interests to declare that are relevant to the content of this article.

\paragraph{Funding.} J\"urgen Lerner was funded by the Deutsche Forschungsgemeinschaft (DFG, German Research Foundation) -- Project number 555455503. Raphael H. Heiberger acknowledges support from the Alexander von Humboldt Foundation. Mark Wittek's research stay at Stanford University in 2024, during which the dataset was curated, was supported by the Academy for Sociology and by the Center for Social and Economic Behavior (C-SEB) at the University of Cologne.

\paragraph{Author contributions.} J\"urgen Lerner: Conceptualization, Methodology, Software, Formal analysis, Writing -- original draft, Writing -- review \& editing. Mark Wittek: Conceptualization, Data curation, Formal analysis, Visualization, Project administration, Writing -- original draft, Writing -- review \& editing. Raphael H. Heiberger: Conceptualization, Data curation, Writing -- original draft, Writing -- review \& editing.

\paragraph{Data and code availability.}\label{para:data}
The data and code required to reproduce the analyses reported in this article are openly available in the project's OSF repository\footnote{\url{https://osf.io/xa2md/}}. The repository contains the analysis data---the covariates for all sampled events, article and author metadata for the oncology subset of the PubMed Knowledge Graph with OpenAlex topic assignments, and the name-based gender classifications---together with the \texttt{R} scripts that estimate the global RHEM (Table~\ref{table:coef_population-level}) and all C-RHEM specifications (Table~\ref{table:AIC_improvements} and Online Resource~1), compute the between-topic heterogeneity statistics (Table~\ref{table:tau}) and the meta-regressions (Figure~\ref{fig:regressions}), and produce Table~\ref{tab:fields} and Figures~\ref{fig:regressions}--\ref{fig:context_moderators}. The covariates were computed with the freely available \texttt{eventnet} software \citep{lerner2023relational}.

\clearpage

\bibliographystyle{abbrvnat}
\bibliography{references}

\clearpage

\appendix

\makeatletter
\renewcommand\@seccntformat[1]{Appendix~\csname the#1\endcsname.\quad}
\makeatother

\section{Details on hyperedge covariates}
\label{sec:hyperedge_covariates}

The network effects used in our analysis are operationalized via hyperedge covariates $x(t,\ell,I,J)$ that have already been formally defined in \citet{lerner2025relational,espinosa2025socio,fabbrucci2025co}. For convenience, we repeat the definitions of the core effects in the following, where we also repeat the qualitative interpretation given in Sect.~\ref{sec:network_effects}.

We recall that data is given as a sequence of scientific publication events
\[
E=\{j_1=(t_1,\ell_1,I_1,J_1),\dots, j_n=(t_n,\ell_n,I_n,J_n)\}\enspace,
\]
where an event $j=(t,\ell,I,J)$ encodes that paper $j\in\mathcal{J}$, having the label $\ell\in\mathcal{L}$ (representing, e.\,g., the scientific topic, area, or topic) is published by the set of authors $I\subseteq\mathcal{I}_t$ at time $t$, citing the set of references $J\subseteq\mathcal{J}_t$.

\emph{Preferential attachment (citation)} is a function of the number of prior citations to papers in $J$. A positive parameter suggests that (sets of) papers with a higher number of prior citations are cited or co-cited with a higher probability in the future, thereby leading to skewed distributions in the number of citations to papers. The precise specification uses the definition of \emph{geometrically-weighted subset repetition} proposed in \citet{fabbrucci2025co}. For a given candidate reference list $J$ at time $t$ we look at all prior publications $j_m=(t_m,\ell_m,I_m,J_m)$ with $t_m<t$, whose reference list $J_m$ overlaps with $J$. Each of these papers is counted with a weight that is geometrically increasing in the size of the overlap $|J_m\cap J|$. The geometric weighting is inspired by geometrically weighted terms used in the specification of exponential random graph models \citep{hunter2006inference}. For a given scaling parameter $\kappa\geq 0$, preferential attachment for citations is defined via
\begin{equation}
  \label{eq:pacite}
\textit{PA cite}^{(\kappa)}(t,\ell,I,J)=\frac{\exp(\kappa)}{|J|}\sum_{t_m<t}\left\{1-\left(1-\frac{1}{\exp(\kappa)}\right)^{|J\cap J_m|}\right\}\cdot|J\cap J_m|\enspace.
\end{equation}
Thus, for each prior event $(t_m,\ell_m,I_m,J_m)$, the fraction of papers in $J$ that are also in $J_m$, that is $\frac{|J\cap J_m|}{|J|}$, is weighted by the factor $1-\left(1-\frac{1}{\exp(\kappa)}\right)^{|J\cap J_m|}$, which converges to one as $|J\cap J_m|$ gets larger. The larger the value of the scaling factor $\kappa$, the slower the weighting factor converges to one. For example, for every positive $\kappa>0$, an overlap of size two with one prior event counts more than an overlap of size one with each of two prior events. Thus, the covariate not only counts the average number of prior citations to papers in $J$, but it additionally counts prior co-citations to pairs or larger subsets of papers within $J$. In our analysis, we set $\kappa=3.0$.

\emph{Preferential attachment (coauthoring)} is a function of the number of prior publications of authors in $I$. A positive parameter suggests that (sets of) authors with a higher number of prior publications publish or co-publish at a higher rate in the future, thereby leading to skewed distributions in the number of authors' publications. The precise definition is almost identical to the one for preferential attachment for citations, with the only difference that this variable considers the overlap of the authors $I$ with the set of authors $I_m$ of prior publications. In particular, for a given scaling parameter $\kappa\geq 0$, preferential attachment for citations is defined via
\begin{equation}
  \label{eq:pacoauth}
\textit{PA coauth}^{(\kappa)}(t,\ell,I,J)=\frac{\exp(\kappa)}{|I|}\sum_{t_m<t}\left\{1-\left(1-\frac{1}{\exp(\kappa)}\right)^{|I\cap I_m|}\right\}\cdot|I\cap I_m|\enspace.
\end{equation}

\emph{Gender homophily (citation)} measures similarity in the ratio of female authors of the citing paper and the cited paper. We first compute the absolute difference of the ratio of females among the citing authors $I$ and the ratio of females among the authors of the cited papers in $J$, which is a measure of gender heterophily. This measure is multiplied with minus one to obtain an indicator of gender homophily. Formally, if $\textit{female}(i)$ is one if author $i$ is female and zero else and if $I(J)$ denotes the union of the author teams of papers in $J$, we define
\begin{equation}
  \label{eq:genhomcit}
\textit{gen hom cite}(t,\ell,I,J)=-\left|\frac{\sum_{i\in I}\textit{female}(i)}{|I|}-\frac{\sum_{i\in I(J)}\textit{female}(i)}{|I(J)|}\right|\enspace.
\end{equation}
 A positive parameter suggests that paper $j$ is more likely to cite paper $j'$ if $j$ and $j'$ have a similar ratio of female authors. In particular, a positive parameter suggests that teams of predominantly female authors tend to cite papers of predominantly female authors and males tend to cite papers of male authors.
 
\emph{Gender homophily (coauthoring)} measures the fraction of pairs of authors $\{i,i'\}\subseteq I$ among the authors in $I$ that have the same gender:
\begin{equation}
  \label{eq:genhomcoauth}
\textit{gen hom coauth}(t,\ell,I,J)=\frac{|\{i,i'\in I\,\colon\; i\neq i'\wedge\textit{female}(i)=\textit{female}(i')\}|}{\binom{|I|}{2}}\enspace.
\end{equation}
 A positive parameter suggests that females typically coauthor with other females and males coauthor with males; from another angle, that teams of coauthors are more homogeneous with respect to gender than expected by chance alone. 
 
\emph{Self citation} measures the ratio of author-reference pairs $(i,j)\in I\times J$ such that $i$ is among the authors of $j$. Formally, if $I(j)$ denotes the set of authors of paper $j$
\begin{equation}
  \label{eq:selfcite}
\textit{self citation}(t,\ell,I,J)=\frac{|\{(i,j)\in I\times J\,\colon\; i\in I(j)\}|}{|I|\cdot|J|}\enspace.
\end{equation}
A positive parameter suggests that authors tend to cite their own prior papers with a higher probability.

\clearpage

\section{Topic-specific estimates}
\label{sec:topic_estimates}

\begin{figure}[h!]
    \centering
    \includegraphics[width=1.0\linewidth]{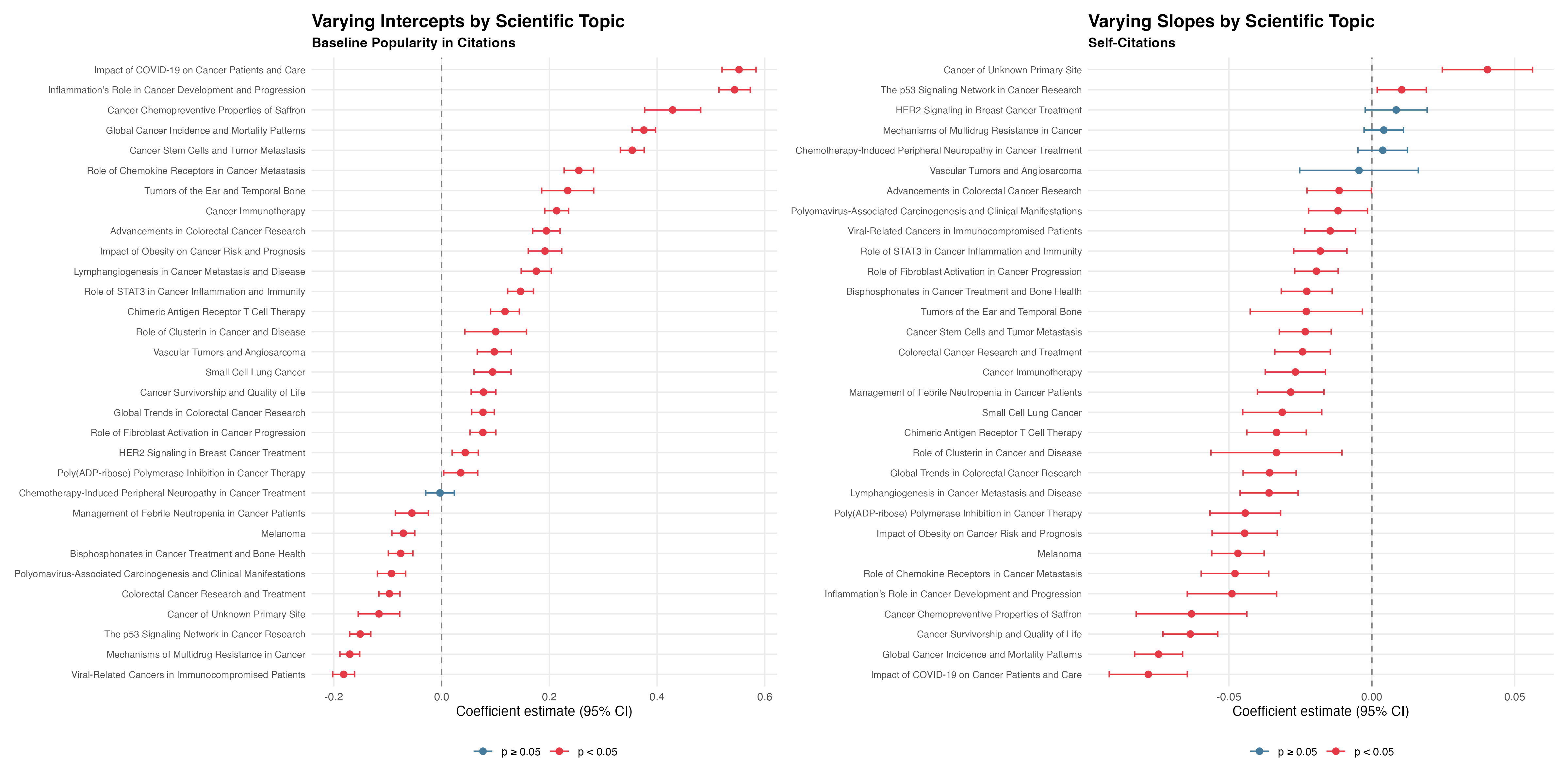}
    \caption{Variation of baseline popularity (left, topic-specific intercepts) and self-citation (right, topic-specific slopes) across topics, relative to the reference topic.}
    \label{fig:base_and_self}
\end{figure}

\begin{figure}[h!]
    \centering
    \includegraphics[width=1.0\linewidth]{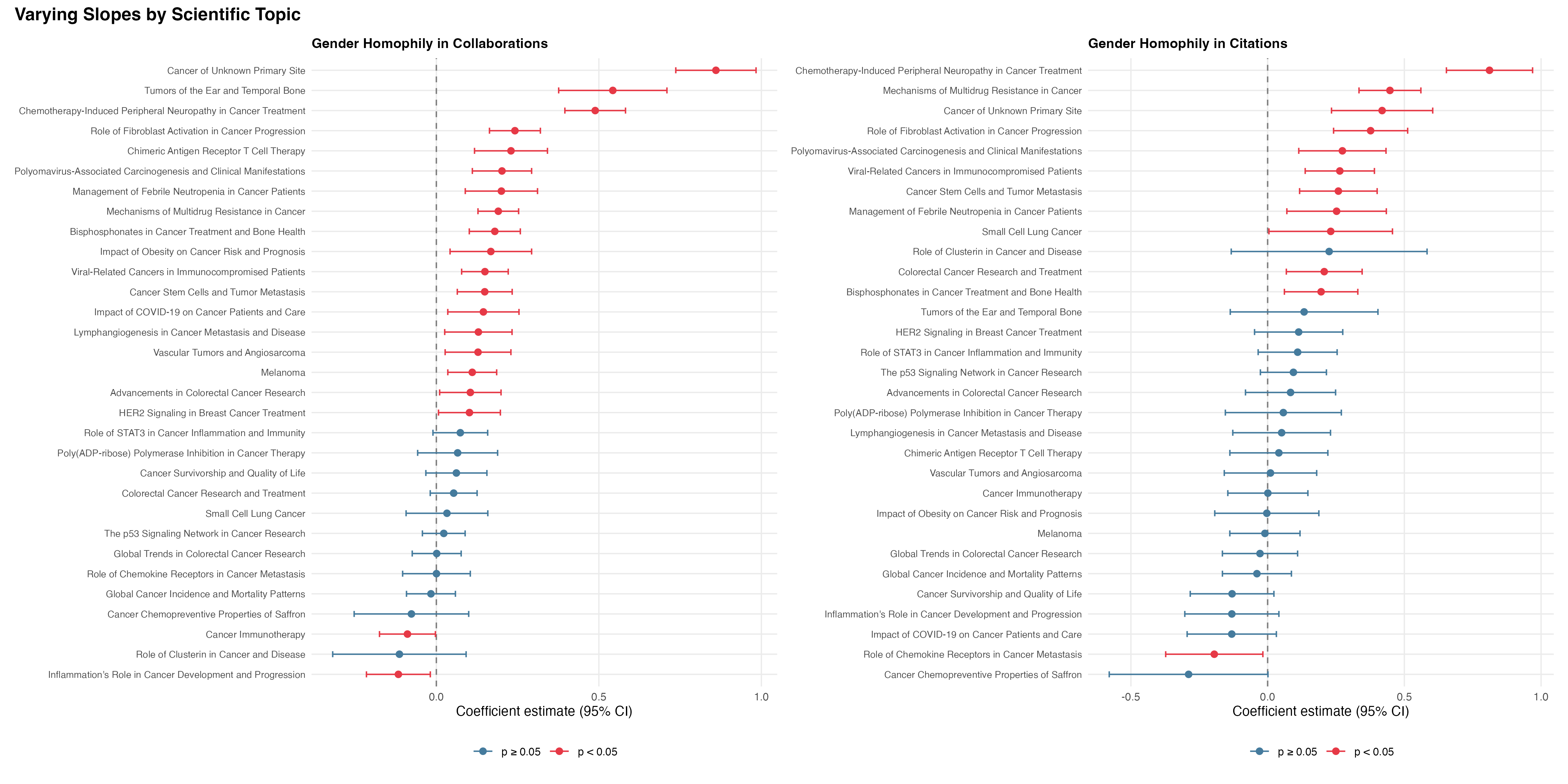}
    \caption{Variation of gender homophily in collaborations and in citations across topics, relative to the reference topic.}
    \label{fig:genderhom_fields}
\end{figure}

\begin{figure}[h!]
    \centering
    \includegraphics[width=1.0\linewidth]{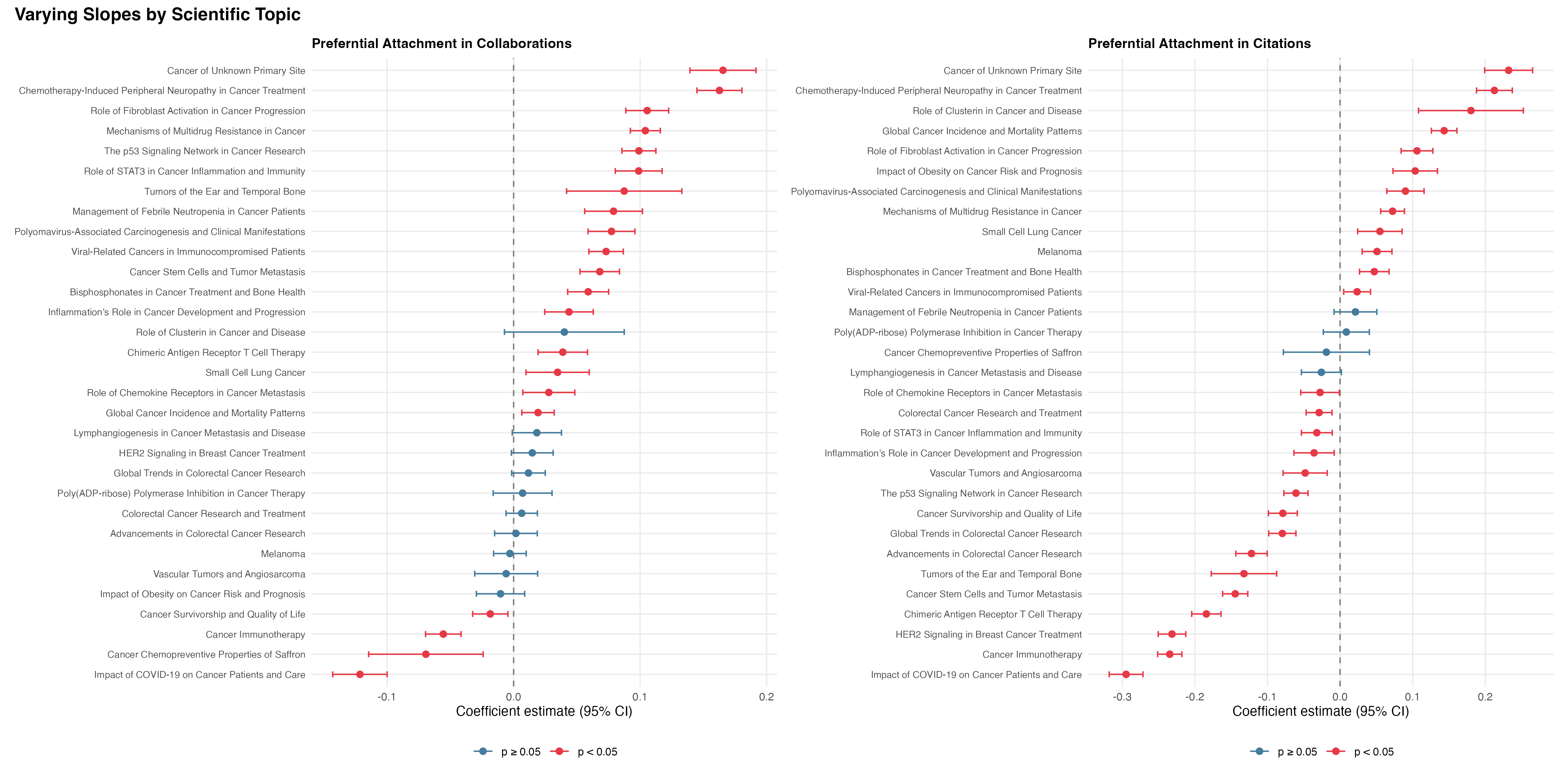}
    \caption{Variation of preferential attachment in collaborations and in citations across topics, relative to the reference topic.}
    \label{fig:prefat_fields}
\end{figure}

Figures~\ref{fig:base_and_self}--\ref{fig:prefat_fields} display the topic-specific intercepts and slopes estimated by the C-RHEMs described in Section~\ref{sec:model_fit}. Estimates are expressed as deviations from the reference topic, \emph{Pancreatic Cancer Research and Treatment} (Topic 10231), the largest topic in our sample; positive (negative) values indicate that an effect is stronger (weaker) in a given topic than in the reference topic. The heterogeneity statistics in Table~\ref{table:tau}, by contrast, are based on absolute topic-specific effects and therefore do not depend on the choice of reference topic.

\emph{Baseline popularity} varies considerably across topics (Figure~\ref{fig:base_and_self}, left). Papers on \emph{Impact of COVID-19 on Cancer Patients and Care} receive substantially more citations than expected relative to the reference topic, whereas papers on \emph{Mechanisms of Multidrug Resistance in Cancer} exhibit significantly lower intrinsic popularity.

\emph{Preferential attachment} (Figure~\ref{fig:prefat_fields}) is positive on average (Table~\ref{table:coef_population-level}), but its strength varies markedly between research areas. Some exhibit highly centralized collaboration and citation dynamics in which productive scientists and highly cited papers accumulate disproportionate attention, whereas others display much weaker cumulative advantage. \emph{Cancer of Unknown Primary Site} shows particularly strong preferential attachment in both collaboration and citation networks, while recently emerging COVID-19-related cancer research exhibits comparatively weak preferential attachment, consistent with the idea that rapidly developing research areas have had less time to establish hierarchical collaboration and citation structures.

\emph{Gender homophily} (Figure~\ref{fig:genderhom_fields}) likewise differs markedly across topics. Several topics exhibit strong gender segregation in both collaboration and citation, whereas others display considerably weaker homophily and thus more integrated interaction patterns. \emph{Cancer of Unknown Primary Site} exhibits among the strongest homophily effects, while \emph{Inflammation's Role in Cancer Development and Progression} displays comparatively weak gender homophily.

Variation in \emph{self-citation} is less pronounced but nonetheless substantial (Figure~\ref{fig:base_and_self}, right). Topics exhibiting stronger preferential attachment often also display higher self-citation rates, whereas recently emerging research areas tend to exhibit weaker self-citation.

\clearpage 

\section{Correlations between topic characteristics and network effects}
\label{sec:descriptives}

\begin{figure}[h!]
    \centering
    \includegraphics[width=0.75\linewidth]{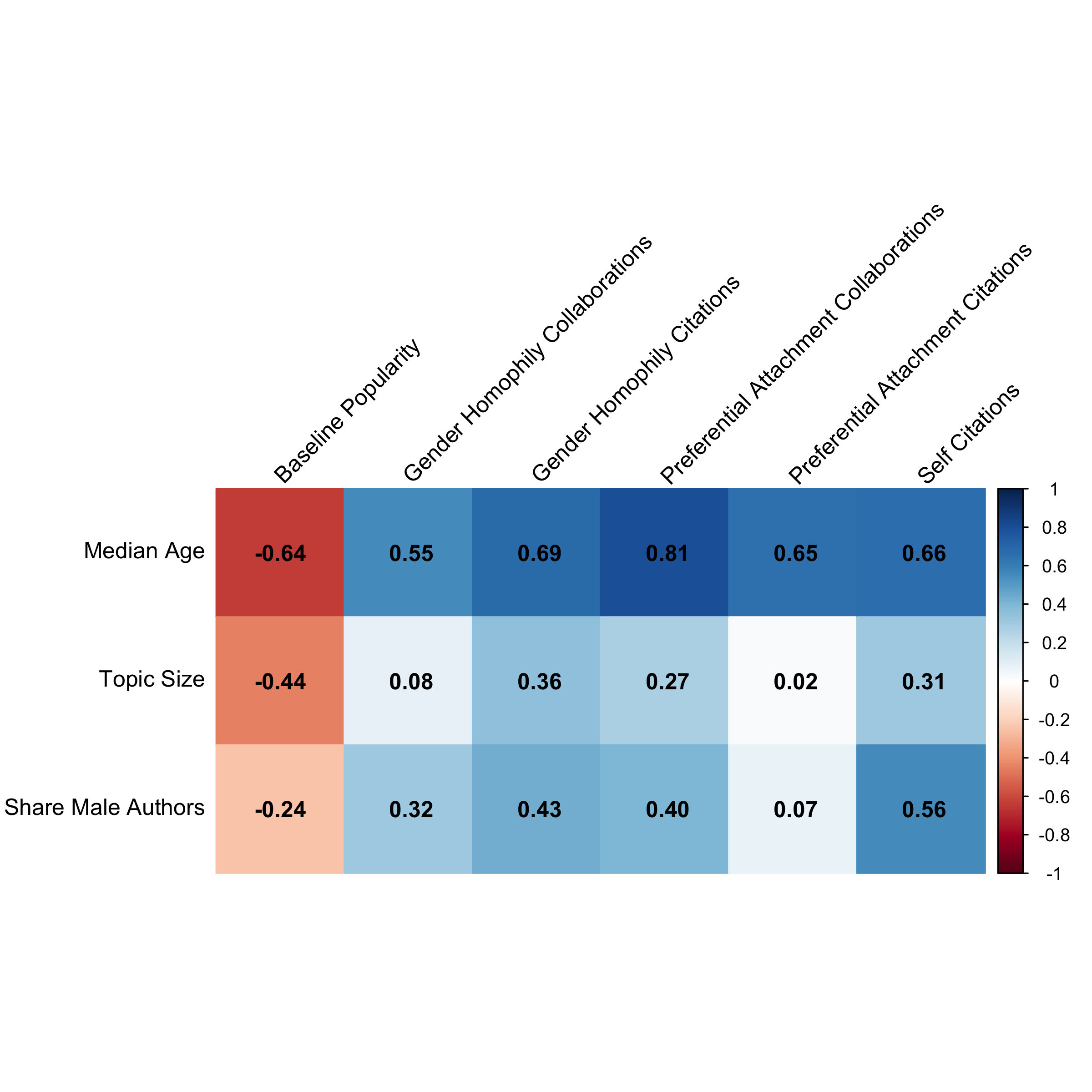}  
    \caption{\textbf{Correlation matrix of topic characteristics and network effects.}}
    \label{fig:context_moderators}
\end{figure}

Figure~\ref{fig:context_moderators} reports bivariate correlations between the three topic characteristics used as moderators in Section~\ref{sec:moderators} and the topic-specific network effects.

\clearpage
\includepdf[pages=-,addtotoc={1,section,1,Online Resource 1,sec:esm}]{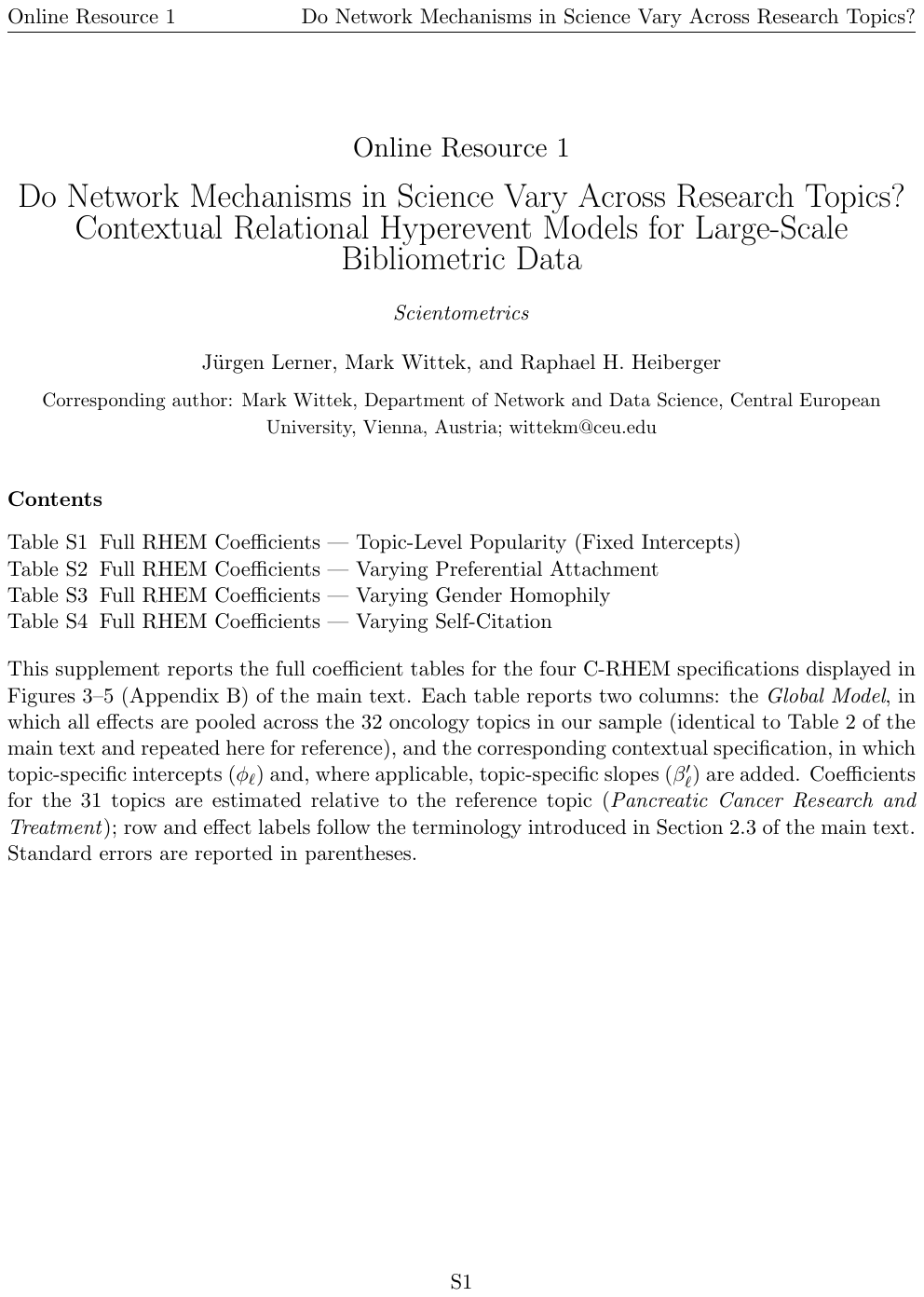}

\end{document}